\documentclass[acmsmall]{acmart}

\usepackage[absolute,showboxes]{textpos}
\usepackage{algorithmic}
\usepackage{graphicx}
\usepackage{subfigure}
\usepackage{textcomp}
\usepackage{standalone}
\usepackage{comment}
\usepackage{booktabs}
\usepackage{balance}
\usepackage{url}
\usepackage{makecell}
\usepackage{tablefootnote}
\usepackage{hyperref}
\usepackage{multirow}

\usepackage{import}
\usepackage{tikz}
\usetikzlibrary{arrows.meta,calc,colorbrewer,decorations,decorations.pathmorphing,decorations.pathreplacing,external,intersections,patterns,shadows,shapes,spy}
\tikzexternaldisable

\makeatletter
\newcommand{\paragraphN}[1]{\vspace*{0.03in}\noindent{\bf #1}\hspace{0.25ex \@plus1ex \@minus.2ex}}
\makeatother

\newenvironment{italicquotes}
{\begin{quote}\itshape}
{\end{quote}}

\newcommand{\supplement}[1]{Supplemental Material [#1]}

\usepackage{todonotes}

\usepackage{pifont}
\newcommand{\cmark}{\ding{51}}%
\newcommand{\xmark}{\ding{56}}%
\newcommand{\bpmark}{\ding{58}}%

\usepackage{xspace}
\newcommand{\etal}{\textit{et al.}}
\newcommand{\eg}{\textit{e.g.,}~}
\newcommand{\ie}{\textit{i.e.,}~}
\newcommand{\etc}{\textit{etc.}~}

\let\orgautoref\autoref
\renewcommand{\autoref}
{\def\sectionautorefname{Section}%
\def\subsectionautorefname{Section}%
\def\subsubsectionautorefname{Section}%
\orgautoref}

\makeatletter
\renewcommand{\paragraph}[1]{\vspace*{0.03in}\noindent{\bf #1.}\hspace{0.25ex \@plus1ex \@minus.2ex}}
\newcommand{\paragraphS}[1]{\vspace*{0.03in}\noindent{\bf #1}\hspace{0.25ex \@plus1ex \@minus.2ex}}
\makeatother

\begin{document}

\thanks{Parts of this work were partially supported by the German Ministry of Education and Research (grant PIVOT).}
\title{Revisiting Randomness in the Constrained IoT: A Systems-level Perspective, Design Guideline, and Brief Survey}
\title{A Guideline on Pseudorandom Number Generation (PRNG) in the IoT}

\author{Peter Kietzmann}
\email{peter.kietzmann@haw-hamburg.de}
\author{Thomas C. Schmidt}
\email{t.schmidt@haw-hamburg.de}
\affiliation{%
  \department{Department Informatik}
  \institution{HAW Hamburg}
  \streetaddress{Berliner Tor 7}
  \city{20099 Hamburg}
  \country{Germany}  
}

\author{Matthias W\"ahlisch}
\email{m.waehlisch@fu-berlin.de}
\affiliation{%
  \department{Institut f\"ur Informatik}
  \institution{Freie Universit{\"a}t Berlin}
  \streetaddress{Takustr. 9}
  \city{14195 Berlin}
  \country{Germany}
}

\begin{abstract}
Random numbers are an essential input to many functions on the Internet of Things (IoT). Common use cases of randomness range from low-level packet transmission to advanced algorithms of artificial intelligence as well as 
security and trust, which heavily rely on unpredictable random sources.  In the constrained IoT, though, unpredictable random sources are a challenging desire due to limited resources, deterministic real-time operations, and frequent lack of a user interface.

In this paper, we revisit the generation of randomness from the perspective of an IoT operating system (OS) that needs to support general purpose or crypto-secure random numbers. 
We analyse the potential attack surface, derive common requirements, and discuss the potentials and shortcomings of 
current IoT OSs. A systematic evaluation of current IoT hardware components and popular software generators based on well-established test suits and on experiments for measuring performance give rise to a set of clear recommendations on how to build such a random subsystem and which generators to use.   
\end{abstract}

\setcopyright{acmlicensed}
\acmJournal{CSUR}
\acmYear{2021} 
\acmVolume{54} 
\acmNumber{6} 
\acmArticle{112} 
\acmArticleSeq{1}
\acmMonth{7} 
\acmPrice{15.00}
\acmDOI{10.1145/3453159}



\maketitle

\setlength{\TPHorizModule}{\textwidth}
\setlength{\TPVertModule}{\paperheight}
\TPMargin{5pt}
\begin{textblock}{1}(.1,0.03)
  \noindent
  \footnotesize
  If you cite this paper, please use the ACM CSUR reference: 
  P. Kietzmann, T. C. Schmidt, M. W\"ahlisch. 2021. A Guideline on Pseudorandom Number Generation (PRNG) in the IoT. \emph{ACM Comput. Surv.} 54, 6, Article 112 (July 2021), 38 pages. 
  \url{https://dl.acm.org/doi/10.1145/3453159}
\end{textblock}

\section{Introduction}\label{sec:intro}

Random numbers are essential in computer systems to enfold versatility and enable security. Almost every operating system (OS) provides ways to generate random numbers. Unfortunately, misconceptions about randomness are common in the design and implementation of operating systems~\cite{cj-rrosh-15}. With this work, we want to shed light on the ever confusing concept of randomness and guide the design of systems that operate constrained embedded devices.

\begin{figure}
  \centering
  \includegraphics[width=0.9\columnwidth]{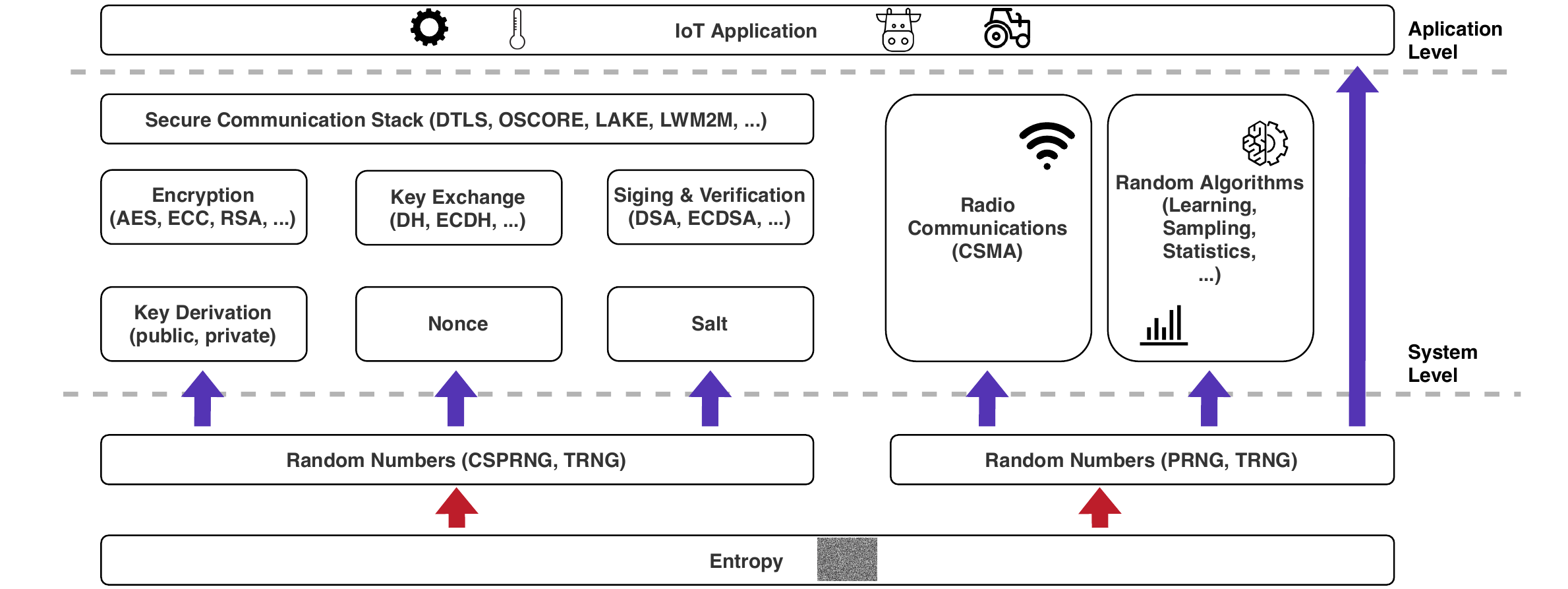}
	\caption{{The role of random number generation in IoT applications.\vspace{-.3cm}}}
  \label{fig:random_application}
\end{figure}

In the Internet of Things (IoT) random numbers are required by various applications and system components, not only in the context of security but also in basic system operations.
Many media access control protocols, for example, apply random delays to avoid interferences.
\autoref{fig:random_application} presents an overview of the typical use cases in an IoT system.

Established IoT applications such as environmental monitoring, automated machinery, machine to machine (M2M) communication, and incident reporting utilize random input.
The need for random numbers increases even further with the advent of Artificial Intelligence~(AI) in the IoT, which shows promise to improve existing use cases as well as to create completely new scenarios on small and cheap devices.
Machine learning at the Internet edge, for example, may be used to pre-process data to improve physical measurements and reduce network load.
This involves randomized algorithms and large-scale random sampling  running on IoT devices.

The Internet of Things (IoT) does not only introduce large quantities of devices to the wild but  in particular resource-constrained, embedded devices (\ie Class~0 to Class~2~\cite{RFC-7228}) to the Internet edge.
IoT nodes of  Class~0 are programmed with bare metal firmware because of extremely limited hardware resources.
Class~1 IoT nodes more and more utilize operating systems instead of bare metal firmware.
More than a dozen open and closed source systems exist that can operate various classes and types of devices. Our work is motivated from the heart of the RIOT system design. RIOT~\cite{bhgws-rotoi-13} contributes randomness to standard networking~\cite{lkhpg-cwemr-18}, security~\cite{fgklp-siici-19}, resource management~\cite{rsw-ehcsp-19}, and other user functions.
Our insights about the creation of random sequences in the IoT can be applied to all types of devices, though.

Producing random sequences on low-end IoT nodes faces particular challenges. As a frequently invoked system service, a random function should be frugal in resources. Neither should it delay algorithms, procedures, or protocols, nor should its state overhead diminish the scarcely available main memory, nor should its energy demands affect the system. Many random algorithms deployed in regular computer systems or clouds are too demanding and violate these constraints. Hence, it is important to identify generators that comply with the IoT resource and energy constraints, which is a major objective of this paper.

Random numbers are also required in almost all security primitives to generate or perpetuate secrets such as encryption keys or cipher streams.
Security protocols and securing communication layers pose strong requirements on the quality of random numbers for guaranteeing confidentiality, integrity, and privacy. The left part of~\autoref{fig:random_application} displays the composition of security components, which are assembled in a crypto-stack. Security protocols (\eg IPsec, DTLS, OSCORE) rely on keys and nonces to encrypt, sign, or verify network packets. Keys and nonces are often generated from random input. Naturally, they must resist prediction and the  random input must consequently be unpredictable, as well. Such demands require a maximized entropy source in the system.
Having a robust security system in place will also enable completely new security applications benefiting from  IoT devices, \eg to implement off-grid blockchain transactions or further variants of distributed ledger technologies.

\noindent In the literature, a ``random sequence'' is often referred to by the definition of \textit{D. H. Lehmer}:
\begin{italicquotes}
``A random sequence is a vague notion [\dots] in which each term is unpredictable to the uninitiated and whose digits pass a certain number of tests traditional with statisticians.''

\hfill \rule[0.5ex]{1.5em}{0.5pt}~\normalfont{Lehmer}~\cite{l-mmlcc-51}
\end{italicquotes}

Ideally, true random number generators (TRNGs) create maximized entropy. True randomness, however, is hard to enforce and  the generation of entropy consumes system resources that are sparse on IoT devices. As an alternative, cryptographically secure PRNGs (CSPRNGs) provide unpredictable random numbers suitable for security purposes. CSPRNGs are based on  a proof that defines hard to solve challenges which an attacker is unable to solve and to break. In order to be fully unpredictable, though, the CSPRNG still requires high entropy for initial seed values.

True random sequences are generated  from random physical
processes such as thermal noise, manufacturing inaccuracies, or crystal drift. Current personal computers
extend such sources to sound or video input, disk drives, user keystrokes,
and more as proposed by the IETF~\cite{RFC-4086}. These interfaces do not necessarily exist on
embedded devices.
There may be complementary sources of random input such as antenna noise or sensor
measurements. Still, collecting random values is difficult and resource-intensive because underlying
processes are slow or do not output continuously.
 Furthermore, running additional hardware components increases energy consumption.

Requirements on the quality of random secrets on constrained devices are similarly strict as with server machines that are easily available to adversaries. Still, cryptographic operations involved in CSPRNGs introduce a notable overhead in processing time and energy consumption on constrained IoT devices that often run on small batteries.
Exploiting this, an adversary who manages to trigger random operations on the IoT device may even run energy depletion attacks. Hence, it is of particular importance to choose algorithms and implementations of CSPRNGs that minimize resource loads while standing up to the standards of unpredictable secrets.

Pseudo-random number generators (PRNGs) are deterministic
algorithms that output sequences with random properties. They require a truly random seed as input in order to start an unpredictable sequence. If properly seeded, these generators expand a comparably short seed
value into a long sequence of random numbers~\cite{dprvw-sapng-13} while discontinuing
 to depend on (physical) random processes. Using a PRNG also reduces the attack surface for adversaries
with physical device access. On low-cost embedded platforms, however, it is hard to generate a qualified seed if  suitable hardware resources are missing. This raises a major challenge for the operating system, as it needs to abstract the hardware without sacrificing functionality.

In this paper, we explore the building blocks for proper  randomness in the constrained Internet of Things.
We start with considering the security impact and the corresponding attack surface (\S~\ref{sec:attack}). Next we discuss the requirements that emerge for  random number generation  (\S~\ref{sec:reqs}) as well as its embedding into a random IoT-OS subsystem (\S ~\ref{sec:environ}). The major contribution of this paper follows with the systematic testing and performance evaluation of common hardware and software generators on IoT platforms. First, we recap the statistical test suites (\S~\ref{sec:eval}) and then analyze popular off-the-shelf  random number hardware along with a seed generator based on PUFs (physically unclonable functions) (\S~\ref{sec:eval_hw}). Second, we perform analogous tests and evaluations for pseudo-random number generating software (\S~\ref{sec:eval_prng}) and deduce recommendations for general purpose and crypto-secure generators in the IoT.
We specifically evaluate randomness in the context of emerging AI applications and hardware platform that represents a device class currently evolving to address AI processing demands at low energy cost on edge devices (\S~\ref{sec:ai}).
Comparing hardware and software performance properties and resource consumption leads to a conclusive discussion on how to combine and jointly deploy the different hardware and software components~(\S~\ref{sec:hwrng_vs_prng}). We summarize four clear recommendations that immediately apply in practice, in the conclusions (\S~\ref{sec:outro}).
The Appendix provides supplementary figures (\S~\ref{appendix:eval_hw}--\ref{appendix:eval_prng}) and lists a table of abbreviations which we use throughout this article~(\S~\ref{sec:appendix}).

\section{The Impact of Random Input on IoT Security}\label{sec:attack}

The Internet of Things extends the distributed Internet system at the edge by a new, massive set of constrained devices. Secure communication in the IoT relies on  cryptographic protocols of consistent design and proper practical instantiation, which includes provisioning of random numbers. Protocol security is commonly built on primitives that can be mathematically proven to meet security requirements. Such proves often rely on the hypothetical presence of a random oracle~\cite{br-eopps-93} that produces truly random input on request. More advanced, complex protocols are then constructed by securely composing these primitives.  Canetti~\cite{c-scmcp-00} was the first to introduce a definition of protocol security that is provably preserved under composition. 

Secrets such as keys or intrinsic function values are spontaneously derived from random input.
 Security degrades whenever randomness is flawed. Extending the desired level of security to the constrained IoT edge is a hard problem and random number generation on embedded devices must be seen as one of the key challenges in this context.
Cryptographically secure random numbers require statistical robustness to withstand statistical attacks. Cryptanalytic attacks, however, encompass additional attack vectors to disclose random state, the previous, and future output. Whenever an attacker can break in the current state of the random system, he should not be able to calculate back previous random values. 
This robustness property is known as forward secrecy and can be achieved by applying a non-invertible  cryptographic function. 

Conversely, the attacker in possession of the current state should neither be able to predict future random output, which is known as backward secrecy. Establishing backward secrecy requires entropy for refreshing the generator state. These operations can be costly on a constrained device and it is the objective of this work to identify feasible solutions of high quality standards. We discuss details of qualitative requirements in Section~\ref{sec:reqs}  and evaluate the trade-offs in Sections~\ref{sec:eval_prng}~and~\ref{sec:hwrng_vs_prng}.

A variety of prominent attacks are based on vulnerabilities of random number generators
in real-world systems and the literature provides a plethora of cryptographic analyses and attack
scenarios~\cite{fc-tcrng-07,nesp-recrt-08,stm-isms-19,kswh-capng-98,ksf-ynday-99,dgp-cwrng-07,ry-wgrgb-10,gpr-alrng-06,dprvw-sapng-13,kkhd-alpaw-14,gw-rnb-96,sf-pbdnd-07,bmss-chpef-10,lhabk-rwwwir-12, tcrh-fibim-20}.
In reality, there are even more attacks anticipated, many of which target at zero day exploits.  Kelsey, Schneier~\etal~\cite{kswh-capng-98} criticize a lack of a widespread
understanding of possible attacks against random number generators among system developers. Given the shortcomings of many random subsystems,
it is worth considering the corresponding attack surface of the specific environment. 

In the following, we present and contrast three common attack taxonomies~\cite{kswh-capng-98,fc-tcrng-07,stm-isms-19} to clarify attacks on the random subsystem of IoT devices: (i) a cryptographic perspective, (ii) an embedded device perspective, and (iii) a systems perspective.

\subsection{Cryptographic Taxonomy}
An attack on a random number generator is an intrusive attempt of distinguishing between the
produced sequence and truly random numbers. This distinction would open doors to predict future
outputs or reproduce recent outputs that might have been used for generation of secrets in the past. The
situation becomes even worse if an adversary manages to direct future numbers of a random
sequence. Kelsey, Schneier~\etal~\cite{kswh-capng-98} enumerate three classes of analytical attacks:

\paragraphS{Direct Cryptanalytic Attacks} monitor PRNG outputs to gain knowledge
about the system in order to distinguish between pseudo-random output and truly random bits.

\paragraphS{Input-Based Attacks} require access to PRNG inputs (seeds and initializations
vectors) to inject known test sequences and perform further cryptoanalysis on random
outputs.

\paragraphS{State Compromise Extension Attacks} base on previously compromised internal state of the
generator and enables prediction or backtracking within the pseudo-random sequence.

These three principle attacks target PRNG core functions and may lead to broken security schemes and protocols.
In the IoT, vulnerable implementations do not only affect privacy concerns \etc but may also lead to actual physical damage because of IoT~actuators.

\subsection{Embedded Device Taxonomy}
Generating random numbers on computers with a human-machine-interface has been studied extensively in the past \cite{bmss-chpef-10,dprvw-sapng-13,dgp-cwrng-07,gpr-alrng-06,kkhd-alpaw-14,kswh-capng-98,ry-wgrgb-10}.
These approaches can also be applied to interconnected embedded devices with our without user interfaces but are exposed to additional attack vectors.
Francillon~\etal~\cite{fc-tcrng-07} introduce two types of attackers for the case of
wireless sensor nodes.

\paragraphS{Remote Attackers} mainly target at cryptanalytic and input-based attacks.
Without accessing the node directly, an adversary tries to compromise or manipulate the
generator state, \eg by monitoring and disturbing communication channels. In this particular
example, wireless noise has been used to generate randomness. An adversary with access to the local wireless network must thus be considered a potential threat to perform an \textit{external attack}.

\paragraphS{Invasive Attackers}  gained read access to the internals of a generator and
 compromised its state at one point in time. This definition does not include write access,
or code injection. Perfect state knowledge of a deterministic pseudo-random
algorithm allows the adversary to predict future outputs and it can reproduce sequences that
have been generated in the past, \ie for cryptographic key generation. Unless true random
vales are added to the generator state periodically, the system remains fully predictable. 
The update interval determines the maximum time that a generator remains vulnerable.
Compromising the state by read access is also named an \textit{internal attack}.

\subsection{System-centric Taxonomy}
In the IoT, a large number of constrained embedded devices inter-connect to each others and to
the global Internet. On the one hand, broadening the networking capabilities increases the
attack interface, especially with availability from the outside of a local network. On the other
hand, simple IoT devices
entail special considerations in comparison to traditional networked devices such as
servers, personal computers, or smartphones. Many IoT devices are very constrained in
hardware capabilities due to minimizing price and form factors, as well as energy resources.
These limitations do not only affect computational power, but often imply sparse hardware protection
features. As a consequence, IoT devices often lack permission management for code execution,
memory protection mechanisms, as well as secret storages.

IoT deployments can grant physical access to the hardware, which
opens a potential interface to analyse and monitor delicate key material, firmware,
or even program execution on a device if tamper detection is not in place. We argue that secure
random number generation
cannot be sustained in the case that an adversary has full read or write access to the device. Shielding
attack vectors without read or write access demands for additional hardware capabilities and manufacturers of
low-power chips already reacted. STMicroelectronics~\cite{stm-isms-19} defines three groups of attacks
against microcontrollers (MCUs).

\paragraphS{System Software Attacks} focus on security and resilience affected by weak implementations, bugs, or insecure
protocols after analyzing or even manipulating program execution. Disturbances are possible even without
device access via network interfaces, \eg by sending malicious packets, or by triggering the execution of non-verified or untrusted library functions that may be
already part of the device firmware. The latter  often relates to ``monkey testing'' or to insider knowledge.

\paragraphS{Hardware Non-invasive Attacks} require hardware access. This category includes any kind
of interface that allows interacting with the device directly, such as debug ports, or bus
interfaces (UART, SPI, I2C, ...). The most dangerous attacks for random sources that
rely on physical processes are based on fault injection. Typically, an adversary exploits the device under environmental
conditions that it is not designed for. Prevalent fault injection parameters are temperature
variations, microwave induction or voltage manipulation. Furthermore, side channel analysis such
as power profiling and timing analyses fall in this category.

\paragraphS{Hardware Invasive Attacks} cover advanced techniques that enable access to
the device silicon with access to hidden secrets, even if  device protection mechanisms are in place. Such attacks are usually very complex and require specific measurement instruments.

Recently, hardware manufacturers of low-power microcontrollers have started to provide different countermeasures to the
physical attack surface, ranging from debug port locks, tamper detection indicators, memory protection units, and
isolated code execution environments. Also, hardware crypto acceleration on the chip becomes more widely available. These features should be
used wherever possible. Nevertheless, many low-cost devices that are supported by
IoT operating systems do not provide all (or any) of these capabilities. Implementations and algorithms used
for random number generation should be designed around the concepts of (i) a high modularity to ease partial use of hardware protection features and (ii) robustness even if hardware protection is missing.

Random number generation cannot be protected, if the adversary has full control over the device.
Analytic attacks as well as fault injections can be shielded, though, by incorporating
cryptographic primitives and carefully gathering entropy for seeding the PRNG, as we will discuss next.

 \section{Generating Randomness in the IoT}\label{sec:reqs}

Every day in the life of an IoT device, random numbers are requested by a variety of use cases. These use cases separate into two classes: either \textit{general purpose} or \textit{cryptographically secure} random input. While general purpose use only requires sufficiently well represented statistical properties, cryptographically secure random numbers must also remain unpredictable even under malicious attacks.  While the first category can be achieved fairly easily, the provisioning of secure randomness is very challenging in the constrained regime and---depending on the attacker model and strength---may not be achievable at all.

\subsection{General Purpose PRNGs}\label{sec:gpprng}

General purpose PRNGs are employed for tasks independent of security aspects. Typical use cases include
the jittering of network protocol timers or media access protocols (\eg random back-off in CSMA) to avoid collisions on a medium.
Other applications of general purpose PRNGs include randomized sampling of sensor measurements and fuzzy testing.

A uniformly distributed stream of statistically independent random numbers is the desired output of a PRNG, which still should be approximated in higher dimensions, since concurrent applications or algorithmic elements may call on sub-sequences of the generator (\eg access every k-th output). Seeds between otherwise identical devices must differ to avoid identical random behavior across devices, and individual seeding after each device restart is desired. Even though seed requirements for general
purposes are moderate, a ``plug and play'' source for gathering seed material is missing on IoT devices that do not provide a
hardware based true random number generator.

General purpose PRNGs are essential on most IoT devices and  frequently called in many use cases. Implementations should therefore be fast and efficient to preserve resources of the constrained devices. Available resources are better spent on
generators with high security demands.

\subsection{Cryptographically Secure PRNGs}\label{sec:csprng}

Crypto-purpose or cryptographically secure PRNGs (CSPRNGs) are generators that are safe to use in security applications, involving the generation of cryptographic keys, nonces, or salts.
Shamir~\cite{s-gcsps-83} introduced the notion of a cryptographically strong PRNG which prevents computation of a desired future output value within certain bounds of time and space complexity. Blum~\etal~\cite{bbs-ssprn-82, bm-htocs-82} introduced cryptographically secure pseudo-random sequences that can be generated in polynomial time, but are unpredictable. Given a preceding output sequence of that generator, but not the seed, it must be computationally infeasible to predict the next bit of output with a better chance than 50\,\%.
Cryptographic system security relies on these random numbers as basic input. Consequently, CSPRNGs are expected to output highly unpredictable number sequences and to be resilient against known attacks.
The security of an implementation goes beyond the scope of computational efforts to predict future outputs and includes countermeasures to protect against weak implementations as well as state compromise by an attacker.

Building a crypto-purpose generator is  more complex and  consumes more system resources than a general
purpose PRNG. It involves additional building blocks of ciphers, cryptographic hash functions,
runtime tests, as well a specifically robust seeding logic.
Computational overhead and especially memory requirements of these building blocks are in potential conflict with resource constraints of IoT nodes, but the availability of secure random numbers is
essential for enabling secure communication over the Internet. In order to reduce software complexity, some microcontrollers provide hardware acceleration of
cryptographic primitives, which should be exploited when implementing the respective components.

A significant body of work reports about failures of PRNGs  and successful attacks against the random input of crypto systems~\cite{kswh-capng-98, fc-tcrng-07,dgp-cwrng-07,ry-wgrgb-10,dprvw-sapng-13,kkhd-alpaw-14,stm-isms-19}.
Hence, provisioning a  cryptographically secure, consistent random infrastructure is a crucial component of a software system, which should
be designed and tested with care. Requirements on CSPRNGs, in-depth analyses of different mechanisms and classification of those
have been presented in~\cite{ks-pfcrn-11,brsns-stsrp-10, bk-rrngu-12,s-cns-14,ry-wgrgb-10,dgp-cwrng-07, RFC-8937}.
We summarize the key aspects of CSPRNGs  in the following paragraphs.

\paragraph{Statistical Randomness}
Any statistical bias gives rise to elementary attack vectors. Even though CSPRNGs mainly consist of deterministic algorithms, a crypto-secure random generator needs to produce sequences that are statistically indistinguishable from truly random~\cite{kswh-capng-98}.
These properties base on the assumption that in a string of (pseudo-random) bits, probabilities for
\textit{one} and \textit{zero} are equal at any time and they are statistically independent.
Even a very small bias
must be considered as potential breach of the randomness assumption and contradicts crypto-requirements of a
secure generator.
A variety of statistical properties can be verified with tests that are available in well
established test suites (see Section~\ref{sec:eval}).

\paragraph{Unpredictability}
CSPRNGs require resistance against external and internal attacks (see Section~\ref{sec:attack}). A common distinction exists between prediction resistance and backtracking resistance. In more detail, prediction resistance means that an
attacker cannot guess  future results in computational time by monitoring the generator history, even
if the algorithm is perfectly known. To achieve this at a given statistical quality, the seed needs to be fully unpredictable. Furthermore,
a crypto-secure PRNGs needs to be built on cryptographic functions, usually one-way hash functions and block ciphers
that are practically not invertible and do not produce colliding output from different inputs.

Every cryptographic system needs to be designed according to a specified security level. An established
threshold is 128 bit of secrecy~\cite{RFC-4086,RFC-6347,RFC-7925}. Assuming an adversary had to guess a secret value, it would require trying about half the number of bit combinations, if all states are equally likely. For 128-bit secrecy this would be $2^{128-1}$ tries
on average to brute force a collision, or $2^{128}$ in the worst case.
This is currently considered secure for computational resources. Both
the seed at the generator input, as well as the internals of its algorithm need to meet the expected security
level. It is important to note that due to the \textit{birthday paradox}, an attack on a cryptographic hash function can
complete with a reduced number of tries~\cite{gcc-gba-88,bk-hfbib-04}. Furthermore, if an attacker gained
knowledge of the internal state
of a generator, it should be ensured that future output is only predictable for a very short time. According to NIST, this should be
achieved by adding fresh and truly random values periodically to the internal generator state  (see Section~\ref{sec:reseed_csprng} on re-seeding).

Backtracking resistance  protects against a reconstruction of previous values or even the seed  after a state compromise. It implies that no correlation between seeds and generated output
should be in place. This behavior is  required to assure perfect forward secrecy within cryptographic protocols~\cite{RFC-2409}.
Backtracking resistance is realized by applying cryptographic functions to the internal state of the
generator and hardened by storing state in protected memory, if available.

\paragraph{High Entropy Seeding}
A truly random seed value is required to make the output of
a CSPRNG unpredictable~\cite{bk-rrngu-12,mu-pcrpt-17}. Random bits are usually extracted from
physical resources and the Shannon entropy~\cite{s-mtc-48} or the Minimum entropy serve as a measure of its randomness. In this context, physically random resources are often referred to as entropy sources. Physical sources of ``randomness''  typically exploit variations in electronic circuits (\eg clock drifts, uninitialized memory, analog-to-digital converter fluctuations), randomly noisy signals (\eg wireless noise, bit errors, thermal noise), or user input signals (\eg key strokes, mouse clicks) which normally are unavailable in the IoT.
These  real-world entropy sources, however, do not always admit ``full randomness'' and additional compression methods are often needed for  maximizing entropy. NIST~\cite{tbkmb-resur-18} also advises that seed generation should not rely on a single entropy source, for resilience.
The IRTF recently proposed methods for improving randomness obtained from weak entropy sources~\cite{RFC-8937}.

Full entropy seeds are required for secrecy in a CSPRNG, which is equivalent to requiring  $2^{n-1}$ tries on average, for guessing  a seed of length $n$ bits. If not seeded with sufficient entropy, an
adversary may  exploit internal state collisions and determine generator output faster. This can drastically degrade the security strength of the generator. Hence, great care must be taken to harvest the number of entropy bits that is required by the  cryptographic strength of the system as defined by and compliant to the algorithm of the CSPRNG~\cite{wd-ansp-19}.
 Caution is advised with implementations that limit the input length of the seeding function.
Fresh entropy may be required repeatedly to
re-seed the internal state of a CSPRNG in order to recover from a potential state compromise, or to
serve multiple generator instances, as we will discuss in the next section.

\paragraph{Health Testing}
The quality of cryptographic system components need particular attention, as it may degrade not only due to software bugs, but also due to hardware aging or side channel attacks. Most entropy sources used for (re-)seeding rely on physical processes and particularly benefit from testing. Self-testing demands increase when physical device access is possible.
A variety of  tests have been proposed by NIST, which should be executed on
all functions of a PRNG and its seeder. These tests range
from known answer testing during validation time up to health tests that are applied during runtime to monitor vitality of entropy sources as well as expected execution of deterministic algorithms. The focus of this contribution is not on testing, and we refer the reader to the specific NIST documents~\cite{bk-rrngu-12,tbkmb-resur-18, bk-rrbgc-16}.

\subsection{A Note on Re-seeding CSPRNGs}\label{sec:reseed_csprng}

A CSPRNG can recover from potential state compromise by regular re-seeding~\cite{kswh-capng-98,ks-pfcrn-11,bk-rrngu-12,wd-ansp-19}.
Re-seeding of PRNGs is often advised~\cite{g-sgpsr-98,fc-tcrng-07,fsk-cedpp-10}
but also under much debate in the literature~\cite{ry-wgrgb-10,dprvw-sapng-13,tcb-eatcw-14}.
Certain
CSPRNGs proposed by NIST even build upon the concept of re-seeding~\cite{bk-rrngu-12}.

Considering that a generator is perfectly secure and was seeded
in agreement with its specified security level, while both its seed and its state are kept in full secrecy, then  an adversary cannot predict the next
output by guessing  within computational time without any re-seeding. So re-seeding becomes unnecessary.
In this ideal scenario, the only reason for re-seeding is to extend the finite period of the specific
pseudo-random algorithm. The period of a generator describes the number of cycles it takes to run through all valid
internal states.  In practice, however, cryptographically secure PRNGs have long enough periods and are not affected by repeating pseudorandom output during their lifetime in a common IoT scenario.
Re-seeding can even be disadvantageous as it introduces an interface to inject
low entropy values to the internal state during runtime. This may foster state collisions and thus break
the resilience against unpredictability.

Entropy is a fragile property that is (i) not always in place (ii) in many cases manipulable at
physical device access and (iii) hard to estimate during runtime. Even worse, it may be hard to depict
failures in case of a compromise. To avoid adding compromised entropy values during runtime, some common security procedures base on a
``trust on first use'' model~\cite{RFC-7435}, which contradicts the re-seeding approach.

Another vulnerability created by re-seeding was revealed by Ristenpart~\etal~\cite{ry-wgrgb-10}. Existing implementations of
entropy collectors cache their outputs in memory pools because---due to its eventually long and indeterministic runtime---entropy gathering is often implemented as a
parallel and asynchronous task. If these
numbers are not consumed immediately, a memory without perfect secrecy  exposes an attack vector. Thus, entropy pools and
internal states should run in trusted execution environments, only. Especially on constrained IoT hardware, this is not
always possible, which makes the case for entropy generation on demand. Conversely, the utilization of insecure memory technologies generally motivates re-seeding with fresh entropy values.

Entropy sources can get compromised during operation,
but the opposite can happen, as well, if sources did not provide full entropy during PRNG instantiation.
In that case, mixing additional entropy values to the internal state during operation can be rescuing.
Finally, few crypto-forums argue that re-seeding protects in case of faulty implementations.
Faulty implementations at hand, however, contradict many assumptions of a cryptographically
secure system.

In summary, there are reasons in favor and against re-seeding of PRNGs and we argue that a decision
should be made by the designer in view of the underlying hardware capabilities, deployment
constraints, application scenarios, and security requirements in place. We conclude that a re-seeding
mechanism should be considered as optional function of a CSPRNG API. This recommendation applies to IoT environments that require modular software in order to adapt to the heterogeneity of
hardware platforms of varying resources and diverse deployments in probably (physically) harsh environments.
These considerations, however, are not limited to resource constrained embedded devices, but apply to regular computers as well.

\subsection{System Components for Generating Randomness}\label{sec:hwsw_classification}

\begin{figure}
    \centering
    \includegraphics[width=0.75\columnwidth]{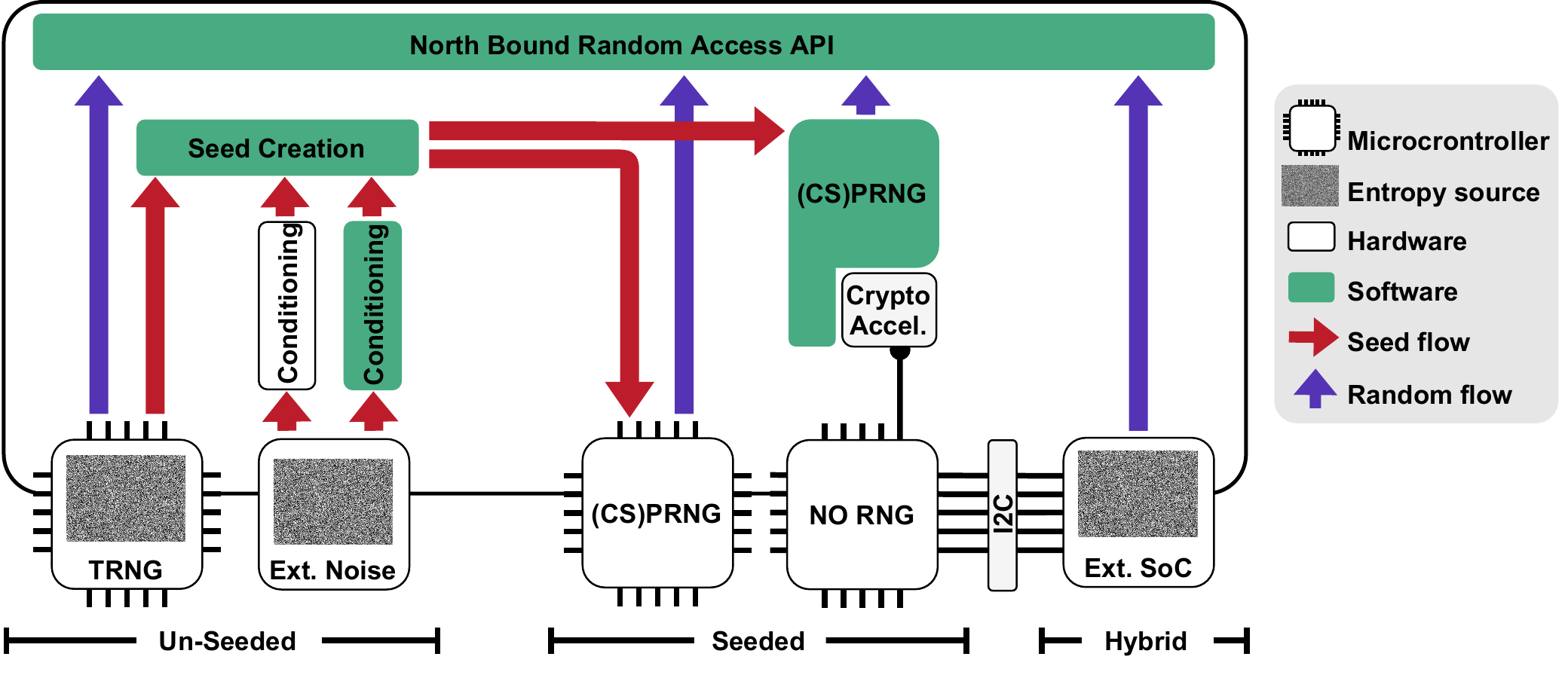}
    \caption{{Overview of hardware and software components for generating randomness in the IoT.}}
    \label{fig:hwsw_classification}
\end{figure}

Random numbers can be produced by hardware or software components. Combinations, in which  assisting hardware improves functionality or performance of a software generator, likewise exist. \autoref{fig:hwsw_classification} presents an overview of different random sources that are commonly available on IoT devices.
The access to the different sources is unified via a ``North Bound Random Access API'', which is commonly provided by the random subsystem of an OS. It is noteworthy, though, that these classifications also apply to ultra-constrained devices which cannot host an operating system. Such bare metal deployments may replace  the random access API dedicated driver or PRNG calls.

We distinguish between (i) unseeded generators, (ii) seeded pseudo-random number generators, and (iii) hybrid solutions.
(i) includes generators of truly random numbers. Many modern microcontrollers provide TRNGs that consist of an internal entropy source and a post-processing hardware circuit that compresses the samples from that source. These sources can feed into the random access API directly, even though it is debatable whether TRNGs should be deployed as an alternative to PRNGs (see~\autoref{sec:hwrng_vs_prng}). Alternatively, external noise from sources such as thermal noise of a resistor, jitter in free running oscillators, or uninitialized memory cells gets sampled. As noisy data provides only few bits of entropy per sample, it needs a separate conditioning,  which can be implemented in hardware~\cite{rv-heppa-19} or software.

The output of TRNGs or (conditioned) noise sources is best used to create start values for seeded PRNGs. (ii) Deterministic general-purpose and crypto-secure PRNGs can be implemented in hardware on the microcontroller itself, or processed in software when random hardware is missing. Software PRNGs can additionally be assisted by hardware acceleration~\cite{kblsw-pscli-21}, which is available for crypto primitives on many platforms.
(iii) Hybrid devices contain an entropy source and pseudo-random hardware, which utilizes the entropy for seeding. This class of devices is composed of dedicated crypto-chips that connect to the main processor using standard communication buses such as I2C, SPI, or UART.

\section{Randomness in IoT Operating Systems}\label{sec:environ}

\subsection{General Requirements}

Many system services require access to random input, and it is common to expect a random function at the operating system level. Use cases and applications of random numbers differ largely, though, as we discussed in Section~\ref{sec:reqs}. General purpose PRNGs are needed to generate random events that follow a uniform statistical distribution and are often consumed at high frequencies.
Security related contexts raise the additional requirement of keeping random output unpredictable, why crypto-secure generators need to maximize entropy with the help of truly random input. Such input is on the one hand difficult to obtain at often high cost, on the other hand truly random sources frequently harvest from system hardware, which is best accessed via the hardware abstraction of an operating system.

Following this perspective, both general purpose and crypto-secure random number generation should be part of an operating system, but are at the same time only versatile if they meet the diverging requirements well. We argue that the different use cases and requirements of PRNGs and CSPRNGs demand for independent methods and APIs. Isolated random functions cannot only be specifically optimized, but also prevent side channel attacks against the CSPRNG via the general purpose PRNG.  Furthermore, separate APIs force developers to decide for their individual use cases.

\subsection{General Purpose PRNGs}

Use cases for general purpose PRNGs require statistically well distributed random sequences.
First, single applications should receive a different value out of the whole number range on each
request to avoid repeating patterns. Second, multiple applications that request from
a single PRNG instance should experience the same properties, even if they access only every k-th PRNG
output. This requires a decent empirical distribution in higher dimensions, which is often a challenge.
Security related applications should not use this generator class as it may be too easy to predict. Furthermore, most go-to PRNGs are invertible which allows to reconstructing previous sequences.
Seeds must be generated differently across devices to prevent a uniform collective behavior---a decent entropy is desired to provide varying sequences between system restarts.
 Seeds may be accessed via the hardware abstraction of the operation system, but should be configurable to ease debugging.

General purpose PRNGs should  be applicable even on very constrained
devices and act frugally while requested frequently. Efficiency metrics involve processing time, as well as energy
 and memory consumption. The latter can benefit from restricting state to a single PRNG instance.
A central instantiation logic can be managed by the operating system.

\subsection{Crypto-secure PRNGs}

\paragraph{Core Requirements}
Security related use cases require crypto-secure PRNGs for sovereign tasks such as cryptographic key
generation. Delicate key material must be largely unpredictable. A
CSPRNG is expected to produce sequences that are indistinguishable from truly random numbers, as discussed in
Section~\ref{sec:csprng}. It is advisable to rely on approved CSPRNG mechanisms that have been
verified by trusted authorities and an operating system or a public library can provide access to
implementations that are tested within this environment. A CSPRNG internally consists of
cryptographic functions~\cite{bk-rrngu-12} to achieve backtracking resistance and the maximum achievable prediction resistance (security strength) is
typically given by that function, though, the strength of the whole generator should be specified by the designer of the approved
algorithm. In order to assure a predefined security strength, a high quality seed must provide
truly random data with a corresponding amount of entropy during instantiation of the CSPRNG.

In contrast to general purpose PRNGs, CSPRNGs undertake tasks like key generation, which is typically involved less
frequently, but also continuous encryption within stream ciphers, thus, performance characteristics of CSPRNGs are important, but secondary in comparison to its security qualifications. Still, computation of
cryptographic primitives and entropy conditioning can be costly~\cite{dprvw-sapng-13}, in particular on constrained
embedded devices. The OS CSPRNG and its seed generator  must comply with the constraints of the
target hardware and leave sufficient resources to deploy a real-world firmware that includes a crypto-stack and the desired application
logic (cf.,~\autoref{sec:intro}).
The operating system should support this in an optimized, configurable environment.

\paragraph{Minimal Standards}
Crypto-secure PRNGs rely on true random seeds that meet a security strength which determines the required amount of entropy in the seed.
At least one entropy source  must be in place that meets the requirements.
The operating system should provide an entropy interface, which grants access to true random
values generated from varying sources, dependent on underlying hardware capabilities.
Externally connected devices may provide true random numbers as well,
but typically require a device driver.  An operating system can simplify access
to relevant components by its hardware abstraction layer, and should additionally allow for code
re-use between different hardware platforms. Configurability requires a highly
modular architecture. In the context of IoT software, configuration is commonly done during
compile time to keep firmware sizes small.

Tests are mandatory for generating robust and secure random numbers~\cite{bk-rrngu-12,tbkmb-resur-18}.
Both the pseudo-random algorithm and the seeding entropy source must be tested, whereas testing procedures can be
separated into a priori and live tests during deployment. Due to device constraints, a priori tests at
development time should be favored  to save resources on running IoT nodes. Thereby, bug free execution of the
deterministic CSPRNG must be verified by comparing output sequences against ground truth. Further, seed sources rely on physical processes and should be evaluated within deployment conditions,
because their behavior can be affected by environmental properties.

\paragraph{Optional Features}
Entropy sources are essential for seed creation, but the properties of underlying  physical processes are diverse.
 Environmental changes as well as attackers with
device access can affect their reliability (see Section~\ref{sec:attack}). Involving multiple
entropy sources during seed creation increases seed resilience.
Naturally, a physical process does not provide full entropy, but conditioning is sometimes
implemented already in hardware on the microcontroller. For sources with sparse entropy concentration, a compression
mechanism is required. An entropy module provided by the operating system can increase seed
quality and it should involve  three fundamental building blocks: (i) An estimate about the
entropy amount per input which can be provided by each source, (ii) an accumulation
instance to involve multiple sources and keeps track of the amount of accumulated entropy, and
(iii) a compression mechanism to create high entropy seeds of limited length to meet security
requirement of the CSPRNG. Steps (ii) and (iii) can be combined in one function. The entropy API should provide  an interface to pass security
requirements and it should also be able to report errors back to the CSPRNG.

Re-seeding a CSPRNG is sometimes desired to recover from potential state
compromise (see Section~\ref{sec:reseed_csprng}). We argue that re-seeding should be kept
optional because seed generation may drain a significant amount of energy on every re-seed
cycle. Enabling and disabling that feature should be transparent to the application that uses
the CSPRNG.

Cryptographic protocols for different purposes require to operate on individual instantiations of a
CSPRNG to prevent side channel attacks. This affects the operating system in two ways. First,
seeding needs to be done separately for each instance---in contrast to the unified approach
suitable for general purpose PRNGs. Second, every instantiation needs its own context to be
handled either internally within the boundaries of the CSPRNG or externally, by dedicating
context allocation to the application.
In both ways, the number of contexts should be kept low in an IoT OS, because the CSPRNG
state can consume much memory. It is worth noting that IoT firmware typically avoids dynamic memory
allocation, why the number of
CSPRNG instantiations should be explicitly defined during development.

The state of a generator needs protection and an operating system should involve hardware
security features, if available. In more detail, secure memory technologies can provide tamper
detection along with authorized access and recent low-power platforms even provide trusted
executions environments
(\eg ARM TrustZone~\cite{arm-trustzone-20})
for protected code execution.
CSPRNG state, seeds, and entropy values should be uninstantiated after use  to avoid
leakage via side channels. This is of particular importance when secure memory is absent.
Memory erasure is commonly done by setting buffers to zero, though, instructions to ``zeroize'' a
buffer are often removed by compiler optimizations, 
which leaves sensitive information in memory.
Known solutions involve \textit{explicit\_bzero} implemented by the GNU C Library, as well as service functions in cryptographic libraries such as libsodium~\cite{libsodium} or Monocypher~\cite{monocypher}.

\paragraph{Optimizing Parameters}
Modern off-the-shelf IoT devices provide on-chip TRNG hardware, which is
commonly used for seeding. In addition,
accelerating hardware units are often in place and capable of processing cryptographic primitives
such as ciphers and hashes. While most CSPRNG implementations base on a pure software solution
of their internal cryptographic functions, a transparent substitution by hardware implementations promises performance
enhancements in terms of speed and energy. 
The operation system can provide peripheral drivers to control hardware accelerators and a transparent reconfiguration that accounts for hardware capabilities. It should automatically
select the most performant solution.

Certain IoT boards provide an external hardware accelerator mounted on the same PCB as the microcontroller, 
which is typically connected by a peripheral bus. Such devices can transparently act as (i) a cryptographic accelerator or
(ii) an alternate CSPRNG, which off-load the main processor while enabling cryptographic applications on limited devices that cannot run CSPRNG software.

\subsection{IoT Operating Systems}

\begin{table*}
\centering
\small
    \caption{Overview of common open-source IoT operating systems and their support for randomness.}
\label{tbl:os_overview}
\setlength{\tabcolsep}{2.8pt}
\begin{tabular}{lllllll}
    \toprule
    & \multicolumn{6}{c}{Operating System}\\
    \cmidrule{2-7}
    & Contiki-NG & mbed OS &FreeRTOS&Zephyr & Mynewt & RIOT\\
    \midrule
    \textbf{PRNG}&&&&&&\\
    \makecell[lt]{\quad General-\\\quad purpose}&
    \makecell[lt]{\textit{rand}\textsuperscript{a}\\(C Library)}&
    Xoroshiro128+&
    \makecell[lt]{\textit{rand}\textsuperscript{a}\\(C Library)}&
    Xoroshiro128+&
    \makecell[tl]{\textit{rand}\textsuperscript{a}\\(C Library)}&
    \makecell[tl]{Mers. Twist.\\Tiny Mers. Twist.\\Xorshift(32)\\Park-Miller LCG\\Knuth LCG}\\
    \makecell[lt]{\quad Crypto-\\\quad  purpose}&
    \xmark&
    \makecell[tl]{HMAC DRBG\\CTR DRBG}&
    \makecell[tl]{HMAC DRBG\\CTR DRBG\\Hash DRBG}&
    \makecell[tl]{HMAC DRBG\\CTR DRBG}&
    \makecell[tl]{HMAC DRBG\\CTR DRBG}&
    \makecell[tl]{HMAC DRBG\\CTR DRBG\\Hash DRBG\\SHA256PRNG\\Fortuna}\\
    \midrule
    \textbf{Entropy}&&&&&&\\
    \quad Sources&TRNG&TRNG&\makecell[tl]{TRNG\\Timer\textsuperscript{b}}&\makecell[tl]{TRNG\\Timer/\\Counter\textsuperscript{b}}&TRNG&\makecell[tl]{TRNG\\SRAM PUF\\CPUID\textsuperscript{b}}\\
    \quad Accumul.&\xmark&\xmark&\cmark&\xmark&\xmark&(\cmark)\\
    \midrule
    \multicolumn{2}{l}{\textbf{Additional features}}&&&&\\
    \quad Ext. state&\xmark&\cmark&\cmark\textsuperscript{c}&\cmark\textsuperscript{c}&\cmark\textsuperscript{c}&\cmark\textsuperscript{c}\\
    \quad Err. interf.&\xmark &\cmark&\cmark\textsuperscript{c}&\cmark\textsuperscript{c}&\cmark\textsuperscript{c}&\cmark\textsuperscript{c}\\
    \quad HW accel.&\xmark&\cmark&\xmark&\xmark&\xmark&\xmark\\
    \bottomrule
\end{tabular}

    \smallskip
    \begin{flushleft}
    \footnotesize{\textsuperscript{a}\textit{rand} maps to an LCG--type PRNG in most libraries.}
    \quad \footnotesize{\textsuperscript{b}Predictable source that varies output between calls or devices.}

	    \footnotesize{\textsuperscript{c}Function or state is not exposed via an OS API.}

    \end{flushleft}
\end{table*}

Currently, the most prominent open source IoT operating systems are Contiki-NG, a successor of the original Contiki operating system~\cite{dgv-clfos-04}, mbed
OS~\cite{arm-mbed-20}, FreeRTOS~\cite{free-rtos-20}, zephyr~\cite{zephyr-20},
Mynewt~\cite{apache-mynewt-20}, 
and RIOT~\cite{bghkl-rosos-18}, all of which implement methods to gather random
numbers. In the following, we
give a brief overview about current solutions in different OSes and summarize the results in~\autoref{tbl:os_overview}. We further on focus on RIOT, which serves as the basis for our experiments.

\paragraphN{Contiki-NG} provides support for a few ARM Cortex-M based microcontrollers, and MSP430 platforms, although its predecessor Contiki provided support for a wider range of architectures. Contiki-NG implements a sparse random subsystem that does not distinguish between general
purpose and cryptographically secure PRNGs. The random API is implemented as a wrapper around peripheral TRNG drivers
of a platform.
The random interface provides a seed
function that limits the input size to an unsigned short integer but it is left unimplemented in most cases,
because the TRNG does not require a seed. In case of missing hardware random number support, the random module falls back to the C library function \textit{rand}.
An entropy module for seed generation is missing.

\paragraphN{mbed OS} is the ARM operating system for processors of its Cortex-M family. mbed implements one  
general purpose PRNG with 
an API interface to re-seed the internal state.
ARM maintains the SSL library mbed TLS~\cite{arm-mbedtls-20} next to the operating system, which generates 
secure random numbers. mbed TLS is portable and used in other  OSs.
The CSPRNG implementations include external state handling, re-seeding procedures, and a collection of self tests.
mbed TLS implements a
dedicated module for entropy gathering, which (i) is capable of accumulating multiple entropy sources, (ii)
provides an interface to add personal entropy
sources, (iii) can be used in a blocking and non-blocking fashion until certain entropy requirements are met,
(iv) distinguishes between weak and 
strong entropy requirements, and (v) can be compiled with different complexity levels.

\paragraphN{FreeRTOS} is a microcontroller OS which is positioned to meet hard real-time requirements. FreeRTOS provides support for a wide range of platforms, including ARM Cortex-A/M devices, RISC-V, and MSP430.
The operating system uses the C library function \textit{rand()} for general-purpose random numbers.
The external mbed TLS and wolfcrypt~\cite{wolfcrpt-20} security libraries
are ported to FreeRTOS, which enable three different crypto-purpose generators. A common random API is missing on the OS level. Self tests, as well as health tests are inherited from the respective library.

\paragraphN{Zephyr} supports a large variety of ARM Cortex-M based 32-bit IoT platforms as well as x86, ESP32, ARC, NIOS II and RISC-V based boards.
The operating system implements a PRNG for general purposes
and the external mbed TLS and tinycrypt~\cite{intel-tinycrypt-20} libraries
are ported to zephyr, which include crypto-purpose generators.
A collection of tests inherited from mbed TLS and tinycrypt can be executed on the operating system, as well as selected benchmarks.

\paragraphN{Mynewt} is an  operating system that supports about 40 boards, most of which with an ARM Cortex-M  32-bit microprocessor, but some also involve MIPS or RISC-V architectures. Similar to
Contiki-NG, it provides access to a TRNG or general purpose random numbers via the C library function \textit{rand()}.
Both  mbed TLS and tinycrypt  are available for cryptographically secure random number generation, but they are not accessible through an OS level random API.
Selftests from mbed TLS can be executed within the operating system, whereas tinycrypt tests are not included.

\paragraphN{RIOT}
currently supports more than 200 boards
involving 30 different microcontroller families that range from 8-bit AVR
devices with sparse peripherals over 16-bit MSP430 devices to 32-bit processors of the ARM Cortex-M family or ESP32  with various MCU peripherals including
dedicated random hardware circuits, as well as ARM7, MIPS, and RISC-V based microcontrollers.
RIOT implements a collection of
PRNGs including ultra-lightweight general purpose algorithms, as well as 
crypto-secure generators. Among other embedded crypto libraries, wolfcrypt, tinycrypt, and relic~\cite{ag-relc-15} run on RIOT, but external
CSPRNGs  are not yet integrated into the random subsystem.
A  \textit{random} API unifies access to pseudo-random numbers, but does not differentiate between general purpose and crypto-secure PRNGs.
A test application with user interaction via the shell allows to evaluate vitality and basic performance metrics of a generator.

\section{Statistical Test Suites for Random Numbers}\label{sec:eval}

The statistical quality of random sequences can be empirically analyzed with  many methods and tools that assess random properties.
The NIST Statistical Test Suite (STS)~\cite{brsns-stsrp-10} and the DIEHARDER Random Number
Test Suite~\cite{beb-drnts-19} combine series of such tests. They are established as standard tools and openly available. Both suites base on hypothesis tests that analyse the input
against the null hypothesis of perfect randomness.  This hypothesis implies that fully deterministic
pseudo-random sequences of ideal random properties cannot be distinguished from truly random values. Conversely, even ideal random sources may produce sequences that appear to have non-random properties, which occasionally leada to failures of statistical tests---usually referred to as type-1 error.

\subsection{NIST Statistical Test Suite}\label{subsec:evalnist}
The NIST STS consists of 15 different test cases, some of which are executed repeatedly, leading to a number of 188 statistics that are processed on each run. With respect to the input size recommendation for each test~\cite{brsns-stsrp-10}, we apply
the test suite version sts-2.1.2 to 100\,Mbit generator output. Every test is repeated 100
times with 1\,Mbit test sequences. A single test returns a probability value
($p$-value) and it is expected to accept the hypothesis of perfect randomness with a confidence of $1-\alpha$, if the value lies above a significance
level of $\alpha=0.01$. Otherwise, the hypothesis of randomness is rejected and the result is interpreted as failure. Each test is applied repeatedly which results in a
vector of $p$-values. The proportion of passed
sequences for one test is determined using the \textit{confidence interval}. As a next step, the distribution of $p-$values for each test is
analyzed using a chi-squared test ($\chi^2$)-test,  which outputs  a second order probability value ($p_2$-value). The test suite defines a significance level of $\alpha_2=0.0001$ for testing this distribution.

\subsection{DIEHARDER Random Number Test Suite}\label{subsec:evaldieharder}
The DIEHARDER test suite subsumes 31 tests to analyse statistics of random input streams. These
tests are executed with varying parameters, thus, a full run calculates a total of 122 test statistics.
We apply the DIEHARDER test suite version 3.31.1 with default options to streams of raw binary
outputs. In default mode, the number of repetitions of a specific test varies
between $psamples$ in [1, 1000] and the sequence length is variable between $tsamples$ in [100, 65000000] which demands for much more random input data in comparison to the NIST STS.
Repeated executions deliver multiple probability values ($p$-values) for each test, similar to the NIST tests.
A Kolmogorov Smirnov test (KS-test) is applied to test deviations from the expected distribution, which results in a second order probability value ($p_2$-value).

In DIEHARDER, a test passes if its $p_2$-value lies above a significance level of $\alpha_2=0.000001$, below
$(1-\alpha_2)$, and it fails otherwise. The result is considered as weak if $p_2$-value lies above
$\alpha_w=0.005$ and below $(1-\alpha_w)$. It is again worth noting that even truly random numbers might generate weak results occasionally.

\subsection{Other Test Suites}

Donald E. Knuth was one of the pioneers who described randomness tests in early editions of \textit{The Art of Computer Programming}~\cite{k-acp-09} and his tests are part of most established test suites today.

NIST released the first random number tests in 1994 within the FIPS 140-1~\cite{nist-srcm-94}
standard, which specified four statistical tests. In 2001, these tests were updated in FIPS 140-2~\cite{nist-srcm-01}
with a narrowing of the test criteria. Both documents served as predecessors for the NIST
Statistical Test Suite (STS) released in
2010, which includes all 140 FIPS  test cases as a subset.
The Diehard test suite was published in 1995 by Marsaglia~\cite{m-dbtr-95}, who had been active
 in this field since years. The test suite implements a collection of 18 test cases,
which are a central part of the  DIEHARDER test
environment, which has been developed since 2003.
Both NIST STS and DIEHARDER are  well known and accepted as standard tools for
statistical testing of random number generators~\cite{es-tclet-07,sz-frtns-14}.

The TestU01 library~\cite{es-tclet-07,es-tslac-13} was introduced in 2007. It includes the
majority of tests from NIST STS, Diehard, as well as additional tests proposed in literature. Its purpose is to provide an ``extensive set of software tools for statistical testing of RNGs.''~\cite{es-tclet-07}, which led to a larger variety of tests, larger sample sizes
and an extended test parametrization in comparison to the other suites. At the core, the environment implements hypothesis tests
similar to NIST STS and DIEHARDER, but instead of rejecting a hypothesis, it simply reports $p$-values outside
the interval [0.001, 0.9990].
TestU01 can
be executed on four complexity levels, of which the most comprehensive one (\textit{BigCrush})
involves up to 160 test statistics. Generation of the required amount of random data can take a long
time, in particular when generated on microcontrollers and transmitted via the UART to feed the library.
This drastically increases time requirements of  the evaluation process.

A range of other test environments are less prominent in the literature.
The SPRNG (Scalable Parallel Random Number Generators)~\cite{ms-asclp-00}
library is a tool to optimize distributed processing for parallel random number generation
and it additionally contributes a few standard tests already covered by NISTS STS and DIEHARDER. The ENT test program~\cite{w-pnstp-08}
defines a small-scale environment that executes only five statistical tests. It relies on a
file based data input, which is not practical when huge datasets have to be analyzed.
The CryptRndTest package~\cite{db-crptc-16} analyses cryptographic random numbers, focusing
on high precision floating-point numbers with lengths larger than 64 bits. The latter is uncommon in the IoT.

\begin{table*}
 \centering
 \small
	\caption{Overview of the typical on- and off-chip IoT hardware with their random features that we analyze. (Abbreviations: RO=Ring Oscillator, LFSR=Linear Feedback Shift Register, RF=Radio Frequency).}
\label{tab:hw_overview}
\begin{tabular}{lllll}
  \toprule
  Board & Chip  & Entropy Source & Post-processing & Error Handling\\
  \midrule
  ST NUCLEO-F410RB	 & STM32F4 	   & 3 Free-running ROs& \makecell[l]{Bias Correction \\ + LFSR}   & Health Tests\\
  Phytec IoT Kit 2 	 &  MKW22D     & 2 Free-running ROs & LFSR& Status Indication\\
  Nordic nRF52840 DK &  nRF52840   & Thermal Noise & Bias Correction & Status Indication \\
  Zolertia RE-Mote   & CC2538  	   & RF Noise (seed only) & 16-Bit HWPRNG& -- \\
  Atmel SAM R21 XPRO & SAMD21  	   & -- & -- & --\\ 
  Arduino Mega 2560  &  MEGA2560 & -- & -- & --\\
  \midrule
  Openlabs Radio Breakout  &  AT86RF233  & RF Noise & -- & --\\
   \vtop{\hbox{\strut Microchip}\hbox{\strut CryptoAuth XPRO-B}} &  \vtop{\hbox{\strut ATECC(5|6)08A}\hbox{\strut }}  & \vtop{\hbox{\strut Quantum Mechanical}\hbox{\strut Circuit Variations}} & \vtop{\hbox{\strut FIPS HWCSPRNG}\hbox{\strut }}& \vtop{\hbox{\strut Health Tests}\hbox{\strut (ATECC608A)}}  \\
  \bottomrule
\end{tabular}
\end{table*}

\section{Hardware Generated Random Numbers}\label{sec:eval_hw}

Common off-the-shelf IoT platforms sometimes provide hardware generated random numbers. While
some platforms implement ``true random''  circuits for entropy gathering on the same chip as the CPU, others implement
pseudo-random generators in hardware. Still, many microcontrollers do not offer random hardware at all.
In these cases, external components such as transceivers or cryptographic co-processors may be
connected to a bus and contribute true random numbers. As an alternative, advanced mechanisms can extract
random physical properties from manufacturing variations of the microcontroller itself.  In
 this section, we analyse typical IoT hardware platforms from different
manufacturers, CPU architectures, and feature sets. Results are summarized in Table~\ref{tab:hw_overview}.
We run RIOT-2020.01 as operating system with a collection of custom measurement programs.

Both the STM32F4~\cite{stm-saabb-18} and MKW22D~\cite{nxp-mrm-16} chips supply a TRNG that gathers entropy from sampling multiple free running and jittering oscillators, followed by a post processor based on a linear shift register that ensures statistically well distributed numbers. They also cover  basic runtime health tests implemented in hardware, as proposed by NIST~\cite{tbkmb-resur-18}.
Although the data sheets claim to pass the NIST statistical test suite, the manufacturer NXP recommends against its direct use for cryptographic applications in place of an approved CSPRNG:
\begin{italicquotes}
``It is important to note there is no known cryptographic proof showing this is a secure
method of generating random data. In fact, there may be an attack against this random
number generator if its output is used directly in a cryptographic application. The attack
is based on the linearity of the internal shift registers. Therefore, it is highly
recommended that this random data produced by this module be used as an entropy
source to provide an input seed to a NIST-approved pseudo-random-number generator
based on DES or SHA-1.''

\hfill \rule[0.5ex]{1.5em}{0.5pt}~\normalfont{NXP}~\cite{nxp-mrm-16}
\end{italicquotes}

The nRF52840 by Nordic~\cite{ns-nps-18} implements a TRNG based on sampled thermal noise followed by an optional post
processor that reduces bias. Even though
the reference manual describes the mechanism as suitable for cryptographic purposes, our results
indicate a slightly different picture without the post processor, as we will show later in this section.

The Texas Instruments CC2538 microcontroller~\cite{ti-ccsoc-13} implements a PRNG in hardware (HWPRNG), which internally consists
of a 16-bit shift register. Hence, its period is limited to  $2^{15}$, in contrast to TRNG peripherals of the previous microcontrollers
The HWPRNG is seeded by noise samples
 on the receive path of the on-chip radio. A similar approach is established on the
standalone AT86RF233 transceiver module~\cite{at-alptf-17}, which produces all random values by
observing noise from the radio.

The ATECC508A~\cite{m-acadc-17} is a feature-rich cryptographic co-processor that runs a NIST-approved CSPRNG (HWCSPRNG)
combined with an internal seed which is inaccessible from the outside. Thus, we consider it as hybrid device (cf.,~\autoref{sec:hwsw_classification}). The
seed is automatically updated on every power or sleep cycle. It can also be updated on demand. 
The seed is generated internally based on entropy extraction from quantum mechanical variations of the
circuitry:

\begin{italicquotes}
``In the crypto devices, the random seed comes from variations at a quantum scale within the
device. The inherent quantum mechanical entropy of the circuitry within the device provides a
truly random seed.''
\hfill \rule[0.5ex]{1.5em}{0.5pt}~\normalfont{Atmel}~\cite{at-grs-15}
\end{italicquotes}

Table~\ref{tab:hw_overview} also lists the SAMD21 and the MEGA2560 microcontrollers by
Atmel as  examples of the numerous off-the-shelf devices, which completely lack
hardware based random sources. This class of devices heavily benefits from external co-processors
as well as internally generated entropy from physical resources that can seed an approved software
PRNG.

\subsection{SRAM PUF Seeder}\label{subsec:puf}

Physically unclonable functions (PUFs) are one solution to generate unpredictable numbers
even without dedicated electronic circuits. They extract unique output from individual hardware properties. Here, we focus on SRAM
PUFs because this memory technology is present on almost all available microcontrollers.
Transistor variations of memory cells lead to varying states after device power-on. The
startup state of multiple memory blocks form a device-unique pattern plus additional
noise, which can be extracted, compressed and used as PRNG seed values~\cite{hlskv-spsco-13,s-lpamb-17,kmg-sswps-17,kgsw-psgri-18}.

\begin{figure}
    \centering
    \includegraphics[width=0.9\columnwidth]{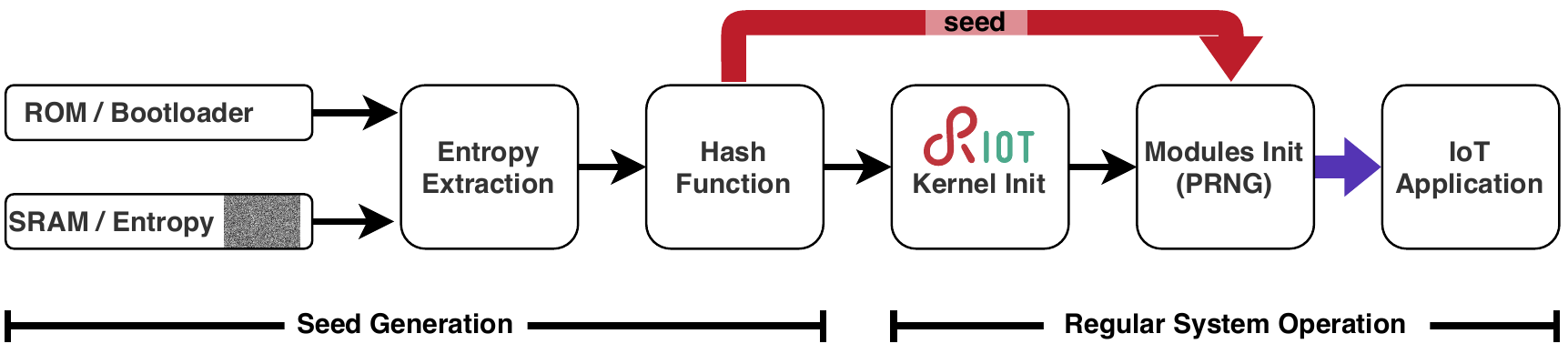}
    \caption{{PUF SRAM random seeder integration in RIOT.}}
    \label{fig:riot-puf}
\end{figure}

\subsubsection{OS Integration}
\hfill\\
\paragraph{Seed Generation}
The mechanism of a PUF-based seeder is visualized in Figure~\ref{fig:riot-puf} for the example of RIOT. It operates
during system startup prior to OS kernel initialization  and reads out uninitialized SRAM cells. A PUF measurement
is compressed by the lightweight DEK hash to build a high entropy 32 bit number that seeds a PRNG during its instantiation,
afterwards in the OS startup sequence. It is noteworthy that the lightweight DEK hash is not a cryptographic function. Furthermore, 32 bit entropy is not sufficient for cryptographic use, but the mechanism is extensible to cryptographic requirements.

\paragraph{Re-seed Power Cycle}
Only startup state of uninitialized memory after re-powering provides high entropy values.
Either a power-off cycle or a low-power cycle without memory retention is required to generate a
new seed. Among others, the required minimum  power-off time depends on ambient conditions, age
and properties of the power source. In our experiments, an off-time of 1\,sec. has proven
suitable for all platforms. Schrijen~\etal~\cite{sl-casmu-12} analyze the impact of environmental and experimental conditions on the SRAM PUF properties in greater detail.

Nevertheless, soft resets or low-power cycles with memory retention can occur and trigger the startup
routine that should not perform a new memory measurement in the absence of a power-off cycle. For a solution, a simple soft-reset detection
mechanism writes a randomly chosen marker at a known and reserved address in the SRAM. During the subsequent
startup, this memory address is inspected and a new memory measurement is only performed if the marker has disappeared. Otherwise, a
soft-reset counter is incremented, added to the previously generated seed and the result is hashed
again for creating a new seed value, which is then stored for the next cycle. The last seed, as well as the soft reset counter could be stored in protected memory for crypto-safe operations.

\subsubsection{Evaluation of the SRAM PUF Seeder}
\hfill\\
Next, we analyse SRAM properties on common off-the-shelf microcontrollers and
present results for the SAMD21 in Tables~\ref{tab:intra} and \ref{tab:inter}.

\begin{table}
    \setlength{\tabcolsep}{3pt} 
    \begin{minipage}[t]{.45\linewidth}
        \centering
        \caption{Min. Entropy and Hamming Weight between 50 reads of 1\,kB SRAM on five SAMD21 MCUs (A--E) at ambient temperature.}
        \label{tab:intra}
        \begin{tabular}{r  r r r r r}
            \toprule
            & \multicolumn{5}{c}{Device}
            \\
            \cmidrule(lr){2-6}
            &\multicolumn{1}{c}{A}& \multicolumn{1}{c}{B}& \multicolumn{1}{c}{C}& \multicolumn{1}{c}{D}& \multicolumn{1}{c}{E} \\
            \midrule
            (i) Entropy [\%] & {4.2\,} & {5.5\,} & {5.3\,} & {4.7\,} & {5.5\,} \\
            (ii) Weight [\%]&50.7\,&  49.5\,&  51.3\,&  49.8\,& 53.1\, \\
            \bottomrule
        \end{tabular}
    \end{minipage}%
    \hspace{.05\linewidth}
    \begin{minipage}[t]{.45\linewidth}
        \caption{Fractional Hamming Distance from 50 reads of 1\,kB SRAM between five SAMD21 MCUs at ambient temperature.}
        \label{tab:inter}
        \centering
        \begin{tabular}{l  c c c c}
            \toprule
            & \multicolumn{4}{c}{Device Pair}
            \\
            \cmidrule(lr){2-5}
            &A--B&  A--C&  A--D&  A--E \\
            \midrule
            (iii) Distance [\%]&49.2\,&   49.8\,&   50.1\,&   50.4\, \\
            \bottomrule
        \end{tabular}
    \end{minipage}
\end{table}

\paragraph{Memory Properties}
As a first step, we inspect the random properties of the memory
in  detail. We analyse intra- and
inter-device variations between multiple PUF responses at
ambient conditions. Therefore, we calculate (i) the minimum entropy as a measure of randomness and
(ii) the hamming weight to determine
bias between multiple startups of one device as well as (iii) the fractional hamming distance
between different device responses to quantify  inter-device uniqueness. Results for (i) and (ii) indicate
existence of a relative min. entropy around 5~\% and unbiased patterns. A relative fractional
distance of approximately 50~\% in (iii) indicates uniqueness of device responses.

\begin{figure}%
    \centering
    \subfigure[Minimum Entropy]{{\includegraphics[width=.5\columnwidth]{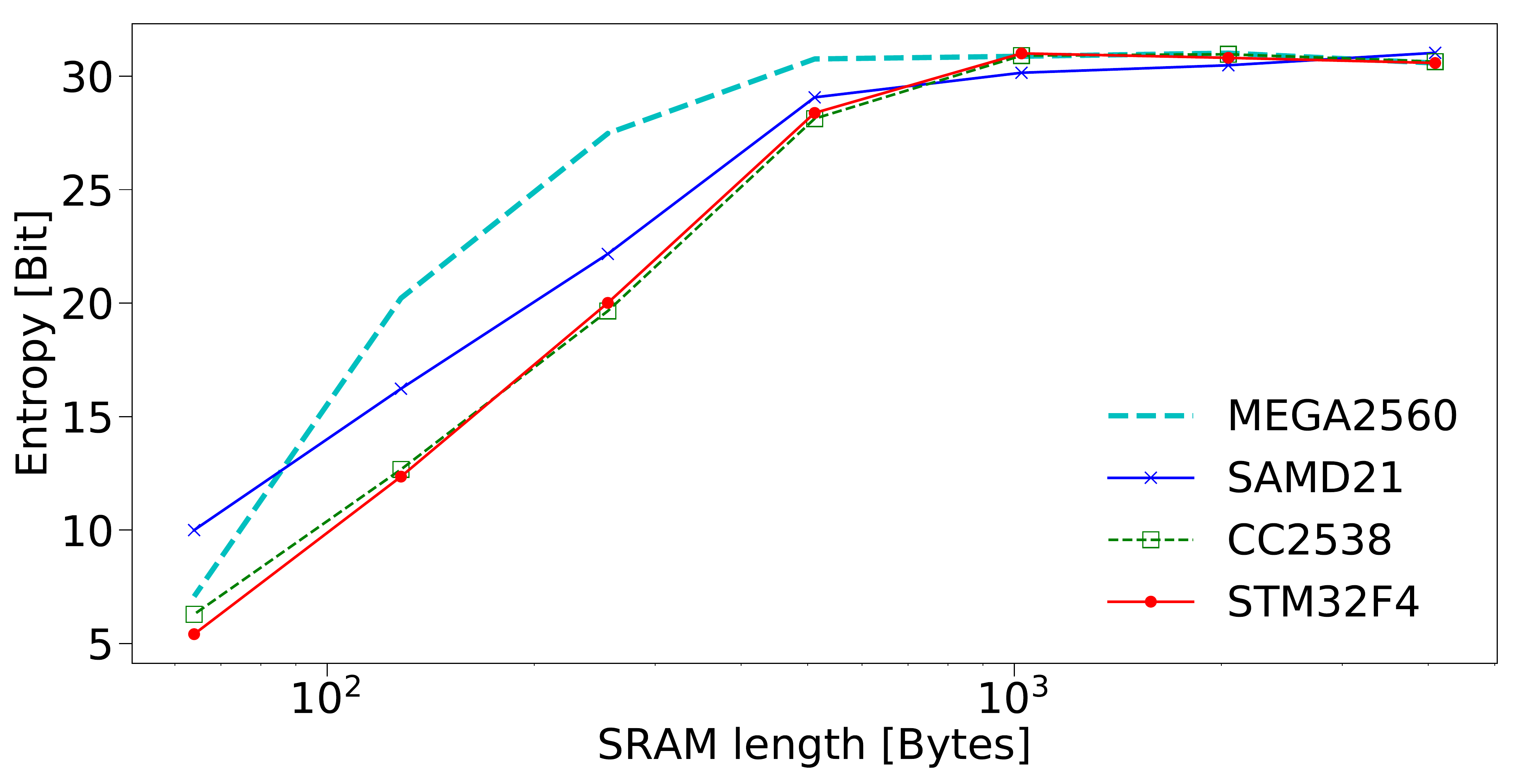} }}%
    \subfigure[Distributions of Hamming Distances]{{\includegraphics[width=.5\columnwidth]{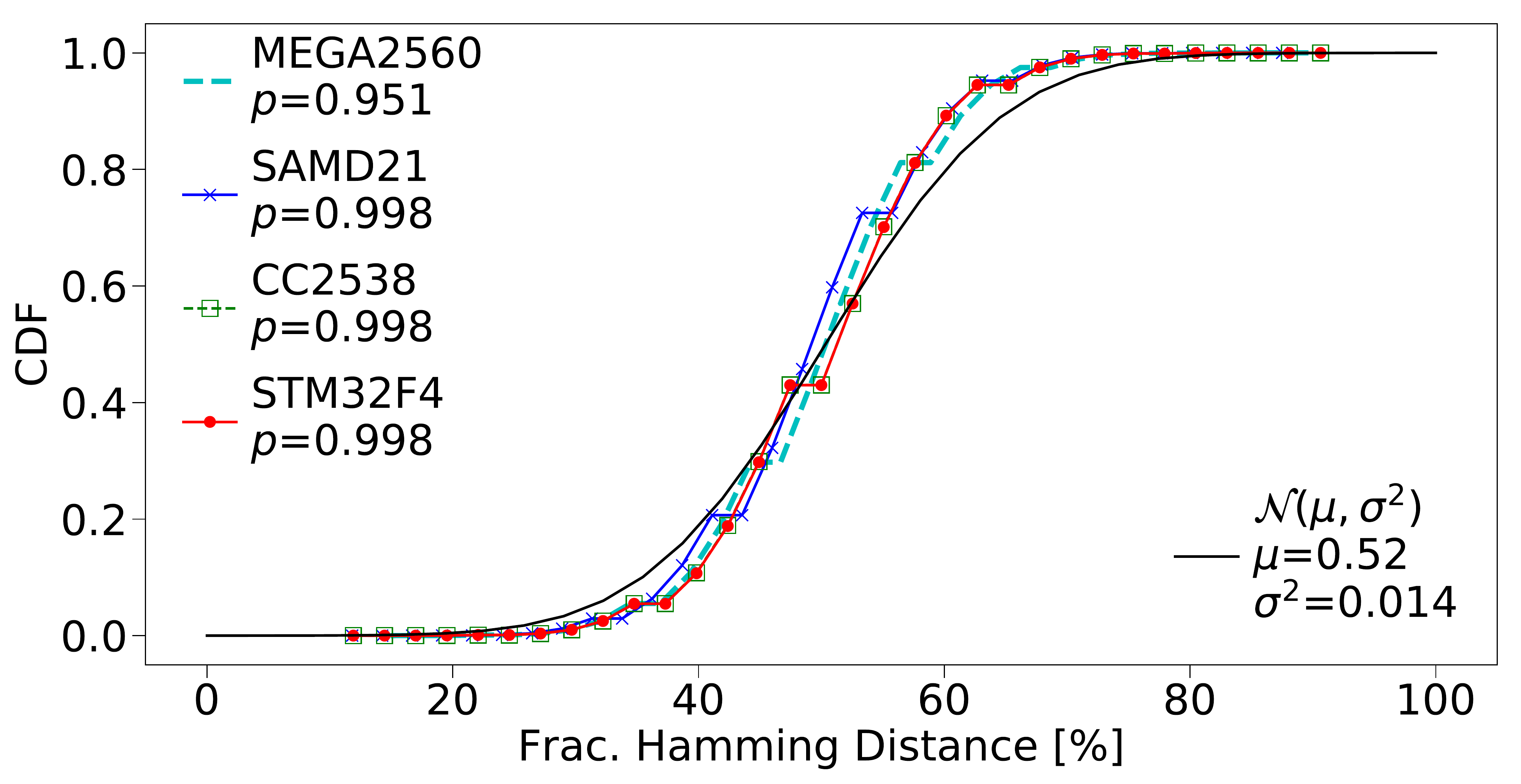} }}%
    \caption{PUF SRAM seed evaluations. Min. Entropy for varying input lengths (left) and distributions of fractional Hamming Distances (right).}%
    \label{fig:seed_probs}%
\end{figure}

\paragraph{Seed Properties}
For determining the proper size of memory used for seed generation we evaluate the minimum seed entropy for
varying input lengths and different platforms as visualized in Figure~\ref{fig:seed_probs} (left). All measurements
converge to approximately 31-Bit entropy with 1\,kB SRAM.

Next, we test the distribution of hamming distances between multiple generated seeds on every device against normal distribution, using a simple
KS-test. The probability function in Figure~\ref{fig:seed_probs} indicates that seed distances on all devices follow a normal distribution with
an average expectation value of 0,52 which is slightly biased by the influence of the MEGA2560 platform. Still, all four controllers pass
the test with probability values greater than 0,95, which allows to accept that they are unique and uncorrelated.

\subsection{Statistical Analysis with NIST STS}\label{subsec:hwsts}

We now evaluate the statistical properties of all hardware based random sources from Table~\ref{tab:hw_overview} by applying the NIST test suite. Following the input size recommendation for each test~\cite{brsns-stsrp-10}, we test 100\,Mbit of random data. Every test is repeated 100 times, which results in 1\,Mbit test sequences. Random integers are generated on the constrained device and fed into the evaluation tool over UART. It is worth noting that already the serial transmission of the data takes at least 45\,min. with a baud rate of 115200. This added to the randomness generation time which led to experiment times ranging from one to over two hours per run. The experiment was executed under office conditions at ambient temperature.

Results of the $\chi^2$-test on the distribution of $p$-values are shown in \supplement{\ref{appendix:hwsts}, Figure~\ref{fig:HW_nist_pval_subplots}}.
All generators but the nRF52840 (w/o bias correction) and the CC2538 show good statistical properties and pass the test suite. The STM32F4 indicates one failure for the Block Frequency test which analyses the proportion of ones in blocks of 128 bit length. A repeated experiment  led to similar results, so we consider this behavior a weakness of that TRNG.

The nRF52840 provides an optional bias correction that is applied to the sampled noise. With enabled correction the nRF52840 (w/ bias correction) passes all statistical tests without deficits. If the post-processor is disabled to increase performance, it fails several of the test statistics. As we will see later, the bias correction requires notably more system resources. We repeated the statistical experiment with and without bias correction multiple times with similar results.

The CC2538 consists of a simple 16-bit linear shift register HWPRNG that (i) shortens the generator period (see Section~\ref{sec:reseed_csprng}) and (ii) can be attacked with sparse processing resources due to the linearity of the internal shift registers. Correspondingly, it fails in most of the tests.

\subsection{Performance Analysis}\label{subsec:hwperform}

\paragraph{Throughput}
We measure the throughput and generation time of different random sources on the STM32F4 platform as well as the internal generators of the
other devices listed in Table~\ref{tab:hw_overview}.
Our test measures the throughput as a single-threaded application that generates streams of pseudo-random numbers continuously, and we count the number of values produced  within an interval of 10 seconds.
In addition, we
measure the time of a single (blocking) function call that returns a random integer with an oscilloscope by toggling an I/O pin via direct register access on the test device. In this way the measurement overhead remains negligible. The ATECC508A crypto-chip always processes 32\,Bytes per request even if only integer values are requested.  We display rates, average processing
times per random integer, and their standard deviation in Table~\ref{tab:hwrng_rates}.

Clearly, the PUF based seeder as well as the externally connected devices perform two up to four orders of magnitudes slower in comparison to the
on-chip generator of the STM32F4 that returns random values via direct register access. It is important to note that our measurements for the
SRAM based seeder only include processing times of the generator itself that consists of the reset detection, memory readout, and hashing to create a final random value. In practice, the required power-off cycle as well as the
OS startup are added, which can take tens to hundreds of milliseconds and is heavily dependent on the OS configuration. However,
in the all-day use of an IoT-device, low power cycles occur repeatedly. The cryptographic co-processor ATECC508A is notably the
slowest but it is the only candidate that runs a hardware based cryptographically secure random number generator. Its
performance strongly relates to the mode of its device driver~\cite{cryptoauthlib}. In our measurements, we use the polling mode, which queries the device for the completed
execution 4--7 times before random data is ready. Alternatively, in the non-polling mode, the driver waits a maximum execution time
for each command, after which the data is ready to be fetched from the co-processor. This potentially increases execution
times but leaves room for parallel processes to be scheduled or low-power sleeping. Furthermore, the ATECC508A as well as the AT86RF233
transceiver require additional I2C/SPI transmissions in comparison to the on-chip solutions. Both the driver overhead and the transport of random data are included in our measurements. It is noteworthy, though, that a real-world deployment of the ATECC508A would process random data internally, which relieves transport over the I/O interface and reduces computation demands of software implementations in a crypto stack.

Among different MCUs on-chip generators, the STM32F4 performs fastest. Although its CPU runs with the highest frequency (96\,MHz),
the processing time is not directly proportional to the CPU speed. In comparison, the MKW22D implements a
similar TRNG and runs at half the CPU speed (48\,MHz), but takes more than ten times longer to produce a random integer value. The
CC2538 unsurprisingly operates comparably fast as it only operates a simple shift register without sampling noise. Both
configurations of the nRF52840 generator indicate low throughput of 6--15\,kB/s in comparison to the other generators, which presumably
relates to the internal sampling procedure. The bias correction leads to statistically good properties of the output
sequences but it reduces computational speed by a factor of 2.5.

\begin{table}[]
    \setlength{\tabcolsep}{4pt} 
	\caption{Hardware Generated Random Numbers: Throughput and processing time per integer (\#).}
	\label{tab:hwrng_rates}
	\centering
	\begin{tabular}{ l  r  r  r | l r r r}
		\toprule
        \makecell[l]{Randomness\\on STM32F4} & \makecell{Rate\\{[}kB/s{]}} & \makecell{Avg. time\\per \# [$\mu$s]} & \makecell{$\sigma$ time\\per \# [$\mu$s]}
        & \makecell[l]{Other\\Platforms} &\makecell{Rate\\{[}kB/s{]}} & \makecell{Avg. time\\per \# [$\mu$s]} & \makecell{$\sigma$ time\\per \# [$\mu$s]}\\
		\midrule
        {STM32F4}  &  &  &
        &{MKW22D} 			& 316 & 33.08 & 0.08\\
        {+ PUF SRAM}  & -- & 296.46\textsuperscript{a}& 0.16
        &{nRF52840}  &  &  &  \\
        {+ ATECC508A} & 3 & 11609.69\textsuperscript{b}& 889.96 
        &{\hspace{3.5mm}w/o correct.} 			& 15 & 246.04 & 2.25\\
        {+ AT86RF233} & 25 & 150.48 & 0.08
        &{\hspace{3.5mm}w/ correct.} 			& 6 & 600.28 & 57.03\\
        {+ TRNG} 			& 1994 & 1.95 & 0.04
        &{CC2538}			& 503 & 6.25 & 0.37\\
		\bottomrule
		\end{tabular}
    	\smallskip
	    \begin{flushleft}
    	\footnotesize{\textsuperscript{a}Does not include time for power-off cycle}
		\quad
		\footnotesize{\textsuperscript{b}ATECC508A always produces 32\,Byte blocks}
	    \end{flushleft}
\end{table}

\paragraph{Energy Consumption}
To evaluate the energy consumption of each generator, we measure the current consumption of all hardware based approaches with a
digital sampling multimeter (Keithley DMM7510 7 1/2) at 1\,MS/s and we drive the board from an external
regulated voltage supply. All development boards provide a measurement header to probe the current that flows to the microcontroller. For
the externally connected devices (\ie ATECC508A and AT86RF233), we additionally measure the current to the power supply pin.
Our energy measurements include driver overhead as well as the transmission over I2C/SPI.
Measurements for the SRAM seeder include reset detection, memory readout, and hashing, referring to our throughput evaluation.
Measurements for on-chip generators include associated testing overheads that are always executed, if available on a hardware platform.
Some hardware based random generators execute at the same scale as the sampling resolution, thus, we measure
generation of 1000 integers per run in that case and normalize the cumulated values afterwards. We repeat every experiment
 1000 times. In setups that involve external hardware, we measure the MCU and the external device separately. If applicable, we separate microcontroller and external device consumption in our graphs.

\begin{figure}
    \centering
    \includegraphics[width=1\columnwidth]{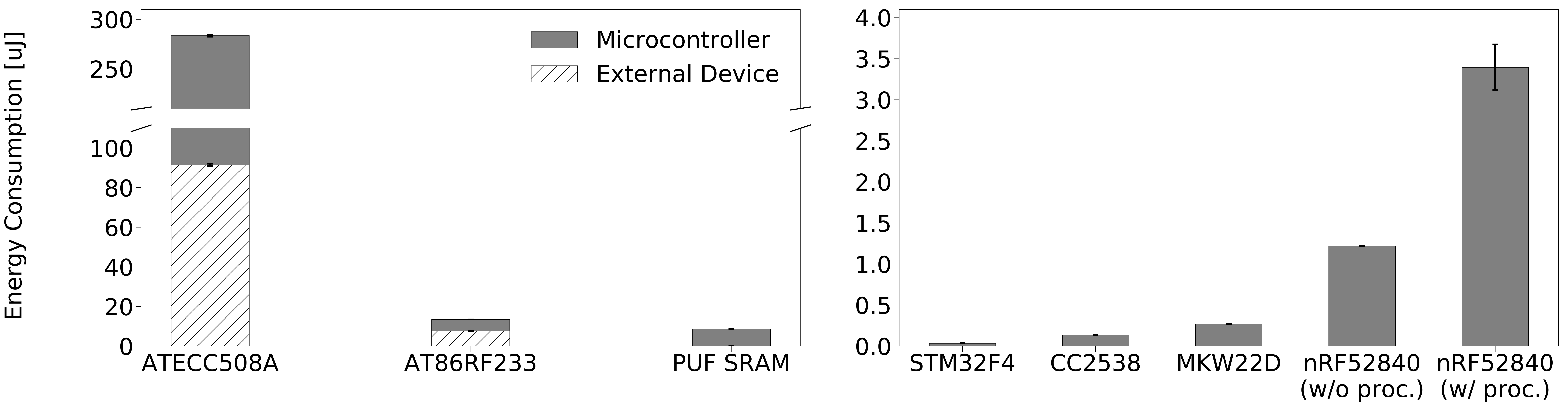}
    \caption{{Hardware Generated Random Numbers: Energy consumption of external (left) and internal on-chip generators (right).}}
    \label{fig:hwrng_energy_bars}
\end{figure}

 Figure~\ref{fig:hwrng_energy_bars} displays our test results on the energy consumption. The radio based approach as well as the SRAM based seeder consume one order of magnitude more energy than the on-chip generated numbers, though, both mechanisms miss an online health testing.
The ATECC508A clearly has the highest energy footprint, which is strongly related to the device driver overhead. As depicted in
Table~\ref{tab:hwrng_rates} the co-processor requires more than 11\,ms to process
the next output, during which the MCU is polling the co-processors status. While co-processing, though,
 the MCU is free to process other data or to go to a deeper sleep mode for energy saving. As depicted in Figure~\ref{fig:hwrng_energy_bars}, the microcontroller consumed the larger portion of energy up to 190\,$\mu$J. Furthermore, the ATECC508A produces a minimum of 32 random bytes per request, thus, up to eight integers can be fetched for the same cost. It is worth noting that continuous health tests are also applied to random samples on the external device.

The radio based random generation is second most energy hungry after the cryptographic co-processor. Although this
approach is ten times faster, it only saves a factor of four in energy,  which is due to the
higher current draw of approximately 12.5\,mA in receive mode of the AT86RF233 during random bit generation. The ATECC508A on the
other hand only drives around 3\,mA during operation. Results for the SRAM based seed generator scale similar to
the radio based approach. It is worth noting that we only display the PUF SRAM overhead here that comes on top
of a power-off cycle. Results for on-chip generators reflect similar properties as the processing times. The consumed energy to produce
one random integer is below 3.5\,uJ on all internal on-chip generators. All devices except the CC2538 platform apply continuous health tests in the hardware. STM32F4 is notably the most frugal competitor and reduces the consumption per integer to
less than 0.35\,uJ.

The difference in energy consumption of the two nRF52840 modes attains about a
factor of 2.5, which increases only slightly less than the processing times listed in Table~\ref{tab:hwrng_rates}.
Still, bias correction introduces a notable increase in processing time and thereby in energy consumption.

\section{Software Generated Pseudo-Random Numbers}\label{sec:eval_prng}

Since von Neumann's early work \cite{n-vtucr-51}, the generic method of obtaining pseudo-random numbers builds up on  some (deterministic) function that is iteratively applied to a (random) seed and attempts to approximate the output of uniformly distributed, independent random trials. As of today, very many of such pseudo-random number generators exist, which comply to various random quality and complexity requirements. From the perspective of an IoT operating system, we are interested in two types of highly efficient, memory-strained algorithms: (i) an ultra-lightweight general purpose PRNG, and (ii) a crypto-secure PRNG that meets the resource constraints of class 1 IoT devices. In the following, we will introduce and analyse eight popular generators---four complex generators of high quality and four lightweight candidates for general purpose random generators in the IoT.

We apply the NIST, DIEHARDER, and the TestU01  suites to these PRNGs using RIOT as OS platform.
Our test programs run as RIOT native processes on Linux servers to speed up experiments.
Thereby, we seeded all
generators  identically, except for the CTR PRNG and the NIST Hash DRBG, which  impose specific requirements on their seeds.

\subsection{Complex Generators}

\paragraph{Fortuna} The Fortuna PRNG is considered a cryptographic random number generator (DRG.4~\cite{ks-pfcrn-11}) and  was designed to
overcome the demand for entropy estimators that tend to be complex and inaccurate~\cite{fsk-cedpp-10}.
Internally, the Fortuna algorithm maintains pools of entropy  from
which a periodic re-seeding is performed. Pools are filled from different available
entropy sources, whose values are distributed among these pools. The entropy accumulation is
conducted by hashing the internal generator state and one entropy pool at a time
using a SHA-256 function. That mechanism allows generating random sequences of unlimited period, though,
it cannot overcome the requirements for proper entropy sources for seeding the generator and for updating entropy pools. Potential
side effects of re-seeding have been discussed in Section~\ref{sec:reseed_csprng}. The final
blocks of pseudo-random output are generated by an AES-128 block cipher in
counter mode.

\paragraph{CTR PRNG} The CTR PRNG is a cryptographic generator (DRG.4~\cite{ks-pfcrn-11}) specified and approved by NIST~\cite{bk-rrngu-12}. It is based on a 128 bit AES block cipher in counter mode, why it provides 128 bit security strength. For this, the generator requires seeds
of at least 128 bits entropy. The tinycrypt library implementation, which we use in our subsequent evaluation,  requires seeds with a minimum length of 256 bits. This is in contrast to all other
implementations presented in this section.

\paragraph{SHA256PRNG} The SHA256PRNG is a generator that provides cryptographically strong random
numbers (DRG.2~\cite{ks-pfcrn-11}). 
The original mechanism of this generator was introduced in FIPS 186--1~\cite{nist-dss-98} and analyzed by Kelsey
~\etal~\cite{kswh-capng-98} and Desaj~\etal~\cite{dhy-ptprn-02}.
Outputs are generated by hashing the internal generator state,  which is updated thereafter by a linear transformation of the hash.

SHA256PRNG is the successor of the SHA1PRNG, which became popular as the choice of
the Java \textit{SecureRandom} class. Until recently, it was considered secure~\cite{g-cidad-02} but got deprecated in
Android N~\cite{android-devs-sha1prng}.
The reasons for deprecation mainly relate to a seeding bug in Java and to NIST deprecating the underlying
SHA-1 hash function because Wang~\etal~\cite{wyy-fcfs-05,sbkam-fcfs-17} discovered an attack that decreased
the number of brute-force tries needed to foster state collisions from $2^{80}$ to $2^{63}$ operations.
For that reason, SHA256PRNG replaces the SHA-1 output function by SHA-256.

The generator is forward secure up to the previous-to-last value, which can be recovered from the current generator output by 
inverting the state update function. SHA256PRNG is confined to a single hash computation per block, which makes it computationally efficient. To add full forward secrecy NIST developed a collection of improved and standardized random bit
generators~\cite{bk-rrngu-12}, the deployment of which has been recommended in revisions of the
FIPS 186~\cite{nist-dss-09,nist-dss-13} standard.

We briefly discuss results about the NIST Hash DRBG (using the SHA-256 hash) at the side. 
It extends the SHA256PRNG by a cryptographic function
that hashes the internal generator state to harden backtracking resistance after state compromise.
This requires two hash computations on each block. 
The NIST Hash DRBG also implements an approved re-seeding mechanism to achieve
unpredictability even after state compromise (DRG.4~\cite{ks-pfcrn-11}).

\paragraph{Mersenne Twister} The Mersenne Twister is a widely used generator, which in its
default version is known to be non-secure~\cite{mn-mtdeu-98}, even though crypto-secure variants have
been explored successively. Known advantages of that algorithm
are a long period and comparably fast operations, because the generation of pseudo-random numbers avoids
multiplication and division. Instead, it requires a large internal buffer of 624 integers.

\subsection{Lightweight Generators}

\paragraph{Tiny Mersenne Twister} The Tiny Mersenne Twister is a derivate of the original
Mersenne Twister that adapts to resource constraints on the price of a narrowed scope. It heavily reduces buffer requirements, on the price of a shortened
period length. The generator is  rather a reduced, memory efficient fallback solution of the full Mersenne Twister~\cite{sm-tmtsv-15}.

\paragraph{Xorshift} The Xorshift generator belongs to the family of linear-feedback shift
register generators that are not cryptographically secure. It is known for its resource efficiency as
it is confined to simple XOR and shift operations. In its simplest 32 bit
state generator, we refer to it in the following. Marsaglia~\cite{m-xr-03} proposed a
collection of extended Xorshift PRNGs with an increased period length and improved statistical
properties. Vigna~\etal\, analyzed and improved these generators further~\cite{v-eemxg-16,v-fsmxg-17,bv-xxgps-18}.
We briefly discuss  results about the Xorshift64* and the Xoroshiro128+ derivatives at the side, which
we integrated into RIOT for  comparison. The Xorshift64* consists of a 64 bit state and applies a
constant multiplication to the output for bit scrambling. The Xoroshiro128+ requires 128 bit state,
although it only outputs 64 bit values per cycle. In contrast to Xorshift64* it adds
 two consecutive state values as a nonlinear transformation to the output.

\paragraph{Park-Miller ``Minimal Standard''} The Minimal Standard algorithm is a linear congruential
generators (LCG) and has been known for decades~\cite{pm-rnggo-88}. Motivated by its
 objective to design a lightweight generator that is confined to
32 bit arithmetic without divisions, it was criticized for its statistical properties repeatedly~\cite{w-pmcpr-05}.
The generator has a period or $2^{31}-1$, and it is limited to produce 31 pseudo-random bits during each cycle.
In RIOT, however, the API presumes 32 bit random integers, thus, one integer is generated by combining  two generator outputs
on each random request, which limits the usable period to $2^{30}-1$.
LCGs are generally not designed for cryptographic purposes.

\paragraph{Knuth LCG} The Knuth LCG is a widely used linear congruential generator, which is
computationally lightweight and has been examined for decades. It is implemented in a range of
software projects and the source code is available in different standard libraries such as
Newlib~\cite{newlib-c-lib} or Musl~C~\cite{musl-c-lib}.
The generator adopts a multiplier that was obtained by Knuth~\cite[Chapter 3.3.2, p. 108]{k-acp-09}
but current implementations differ from the ``MMIX'' Knuth LCG
in its increment. Furthermore, it truncates the most significant
bits of the 64 bit state to 32 bit output values due to known poor statistical properties of
the lower bits in  modulo-2 generators~\cite{m-rnfmp-68}.

\subsection{Statistical Analysis with NIST STS}

Our first statistical analysis applies the NIST STS test suite to these generators.
The results of the $\chi^2$-test are shown in \supplement{\ref{appendix:prngsts}, Figure~\ref{fig:nist_pval_subplots}}. 
Analog to Section~\ref{subsec:hwsts}, tests are only passed after a significant
proportion of repeated successful runs. That is, 96 successful results for a sample size of 100 sequences.
Test 11 and Test 12 (Random Excursions and Random Excursions Variant), however, are not always applicable and as such they reduce the sample size internally.
In our configuration, this led to proportions of sequences passing one of both tests in ranges from  53/56 to 67/71 in different PRNG measurements.

All generators except for the Xorshift pass all 15 statistical NIST  tests.
Xorshift flaws in the Rank (Binary Matrix Rank) test.
This test analyses  linear dependences among substrings of a sequence. In our
case, all 100 test sequences fail already the first order hypothesis test and thus, the $\chi^2$-distributions test fails in consequence.
We consider this an algorithmik weakness. The deficit is fixed by the extensions applied to Xorshift64* and Xoroshiro128+, which pass all tests.

The Frequency Test on the Mersenne Twister had a very small $p$-value on the first run (not displayed here) and we ran the same test again with altered seed value during initialization. 
We want to stress that the magnitude of a $p$-value in hypothesis testing is not
 a causal measure of quality. This means that a low $p$-value confirms to reject a null hypothesis (sequence is not random) but it does not
claim how likely it is that the alternative hypothesis is true (sequence is random). For further interpretation of statistical tests, we refer the
reader to the NIST test suite description~\cite{brsns-stsrp-10} and work by Greenland~\etal~\cite{gsrcp-at-16}.

\subsection{Statistical Analysis with DIEHARDER}
Next, we apply the DIEHARDER test suite to our random generator candidates. DIEHARDER test values behave similar to the NIST STS results presented in the previous section, but procedures are more demanding and failures are
 more distinctive.

Results of the Kolmogorov-Smirnov test for each PRNG in RIOT are displayed
in \supplement{\ref{appendix:prngdie}, Figure~\ref{fig:die_pval_subplots}}. 
All complex generators (\ie Fortuna, CTR PRNG, SHA256PRNG, and Mersenne Twister) pass all tests.
One test returns weak results, which is expected in 1~\% of the test cases due to the uniform distribution of p-values. The DIEHARDER help page~\cite{dieharder-linux-manpage} recommends repeated test
executions and analysis of p-value histograms on weak results. All weak results 
passed a repeated experiment run with a different seed.

Several failures must be observed for the lightweight generators (\ie Tiny
Mersenne Twister, Xorshift, Minimal Standard, Knuth LCG).
The Tiny Mersenne Twister fails Tests OQSO (Overlapping Quadruples Sparse Occupancy Test), and
the DNA test, which both examine the distribution of overlapping substrings in a stream of
random integer values. These results indicate a systematic problem of the generator.
The Xorshift generator fails the Monobit 2 test, the 32x32 Binary Rank test, and the Count the 1s Stream test.
The Monobit 2  is a derivate of the NIST Frequency test and measures the proportion of zeros and ones
within blocks (12-bit blocks applied here). Surprisingly, the general Monobit test, which counts ``1'' bits in a long sequence of random samples
(100000 samples considered here) does not fail. Hence, the Xorshift introduces bias within smaller sub-blocks, which is
compensated over the whole sequence. Results persist after repeated experiment executions with different seeds.
Failing on the matrix
rank test was expected, as the equivalent from the NIST STS already failed  in a similar configuration. The Count the
1s Stream Test examines whether the distribution of ones in a stream of bytes matches that of uniform random bytes
(Binomial(8, 0.5)). Failures indicate that Xorshift output streams produce repeated ``words'' that appear with
pronounced probability. It is likely that the same effect led to bad results of the linear dependency test among sub-matrices.
The advanced Xorshift64* and Xoroshiro128+ generators pass all DIEHARDER tests.

The Minimal Standard generator fails the Bitstream and the Generalized Minimum Distance test. The first
successively analyses overlapping 20-bit tuples ($2^{20}$ possible words) and tests the statistic of missing words for a  normal distribution. Failing this test indicates recurrence of patterns with enhanced probability. The second test places random pairs of points in a square and tests its squared distances for an exponential distribution. The Minimal Standard generator fails to produce outputs that appear independent in this dimension.
Finally, this generator issues a suspiciously high number of weak test results.
For further interpretation of test results, we refer to the DIEHARDER Test Suite description~\cite{beb-drnts-19}.

\subsection{Statistical Analysis with TestU01}

We additionally apply the ``BigCrush'' from the TestU01 test suite to all software generators.
Table~\ref{tab:testu01}  summarizes test results with failures reflecting the number of reported test statistics with $p$-values outside the confidence interval [0.001, 0.9990].

\begin{table*}
    \small
    \caption{Summary of test results from the ``BigCrush'' of the TestU01  environment.}
    \label{tab:testu01}
    \centering
    \begin{tabular}{lcccccccc}
        \toprule
        \makecell{Generator\\ \quad} & \makecell{Fortuna\\ \quad}&\makecell{CTR\\PRNG}&\makecell{SHA256\\PRNG}&\makecell{Mersenne\\Twister} & \makecell{Tiny Mers.\\Twist.}&\makecell{Xorshift\\ \quad}&\makecell{Minimal\\Standard} &\makecell{Knuth\\LCG}\\
        \midrule
        Failures&1/160&0/160&0/160&2/160&13/160&58/160&71/160&9/160\\
        \bottomrule
    \end{tabular}
\end{table*}

Results reflect a similar picture as depicted by NIST and DIEHARDER. TestU01, though,  stresses more failures due to a
higher number of tests and tighter hypothesis criteria. According to L'Ecuyer~\etal~\cite{es-tclet-07}, $p$-values outside the significance interval are obtained approximately 2\,\% of the time, even if the PRNG behaves well. This, however, should not reoccur systematically.

The Minimal Standard misses almost half of all tests (45\,\%), and the Xorshift fails a
surprisingly high number of 58 tests (35\,\%),  which  reduces to 1--2 failures for its enhanced derivatives Xorshift64* and Xoroshiro128+. The Tiny Mersenne Twisters attains a failure rate of 8\,\%, which  still exceeds the acceptable rate by a factor of four, whereas the Knuth LCG performs notably better, failing about 5\,\% of
the test statistics.

For the Mersenne Twister, both failures have a $p$-value of more than $(1-10^{-15})$, which significantly
 misses the acceptance interval, and unacceptable results reappear in repeated experiments with different start values.
This indicates systematic weaknesses. In contrast, all CSPRNGs report zero or singular failures based
on $p$-values at the order of $10^{-4}$, which disappear for repeated tests.

\subsection{Performance Analysis}\label{subsec:eval_prng_performance}

In the constrained IoT, an important dimension for any base system primitive lies in its performance. To assess the value of the different random number generators, we measure the computational speed, the memory overhead, and the energy consumption of all pseudo-random number generators. Thereby, we consider base mechanisms, and we disable re-seeding, if available.

\paragraph{Throughput}
We measure the throughput and speed of each generator on the STM32F4 hardware platform that has been introduced in Section~\ref{sec:eval_hw}.
Our test applications are implemented like presented in Section~\ref{subsec:hwperform}. In addition, we
display maximum values for cases, in which a generator occasionally takes significant time to rebuild its internal state.
Results are summarized in Table~\ref{tab:prng_rates}.

\begin{table}[t]
  \setlength{\tabcolsep}{3pt} 
  \begin{minipage}[]{.5\linewidth}
    \flushleft
    \caption{PRNG throughput and processing time per integer (\#) measured on STM32F4.}
    \label{tab:prng_rates}
    \begin{tabular}{ l  r  r  r}
      \toprule
      Generator & \makecell{Rate\\{[}kB/s{]}} & \makecell{Avg. time\\per \# {[}$\mu$s{]}} & \makecell{Max. time\\per \# {[}$\mu$s{]}}\\
      \midrule
      {Fortuna} & 44 & 87.50 & --\\
      {CTR PRNG} & 102 & 442.01\textsuperscript{a}& --\\
      {SHA256PRNG} & 393 & 10.04 & 69.60\\
      {Mersenne Twister} & 3605 & 0.85 & 189.20\\
      {Tiny Mers. Twist.} & 4807 & 0.62 & --\\
      {Xorshift} & 8152 & 0.28 & --\\
      {Minimal Standard} & 3348 & 0.98 & --\\ 
      {Knuth LCG} &6147 & 0.44 & --\\
      \bottomrule
    \end{tabular}
    \smallskip
    \footnotesize{\textsuperscript{a}CTR PRNG calculates at least one AES-128 cipher per call}
  \end{minipage}%
  \hfill
  \begin{minipage}[]{.45\linewidth}
    \includegraphics[width=1\linewidth]{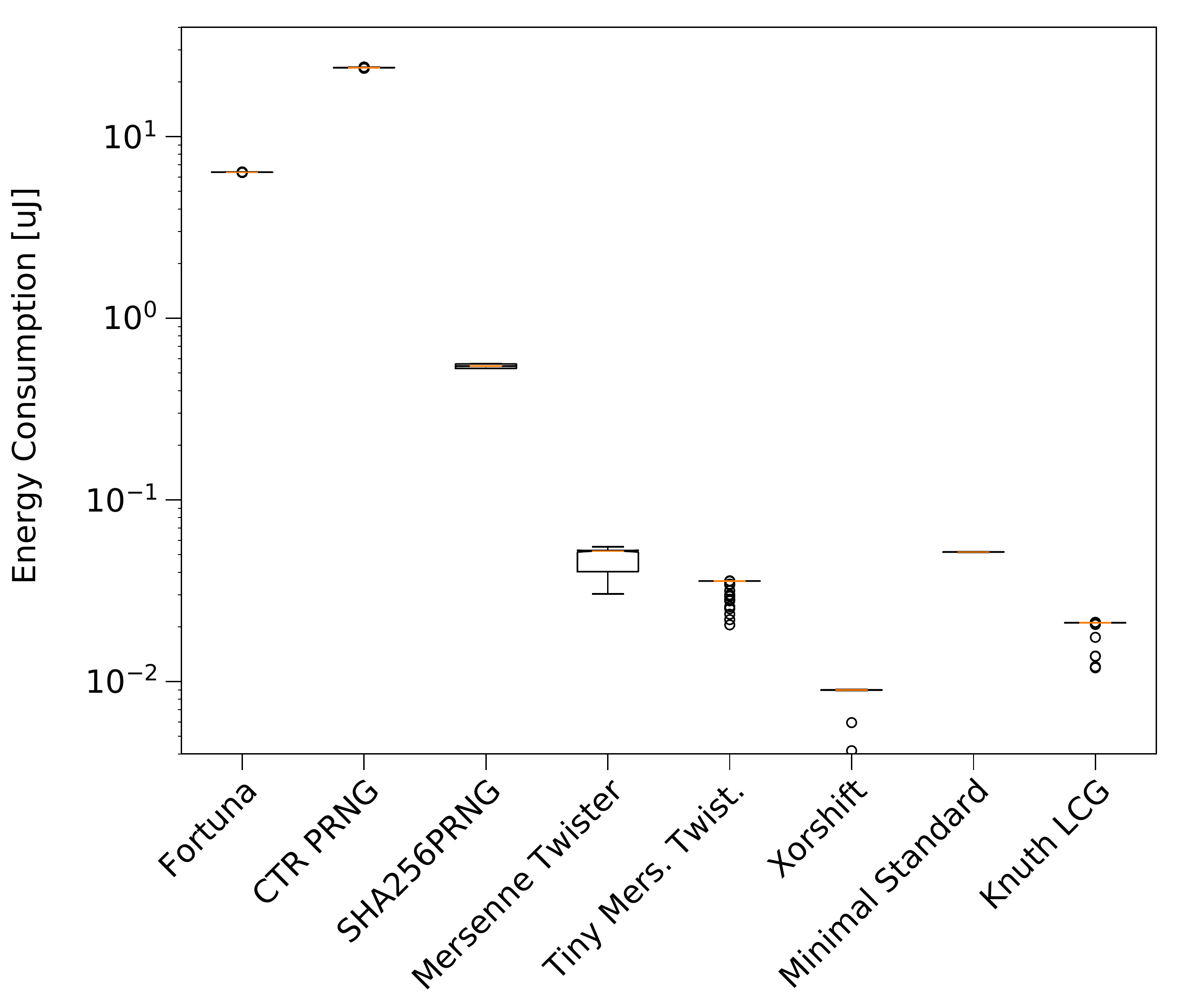}
    \captionof{figure}{{PRNG energy consumption per integer measured on a STM32F4 microcontroller.}}
    \label{fig:prng_energy_boxes}
  \end{minipage}
\end{table}

Naturally, the four lightweight generators are fastest. They can reliably compute a random number in less than a microsecond. The two ultra-lightweight algorithms Knuth LCG and Xorshift can even produce several numbers per microsecond on the constrained microcontroller, which must be considered a very low runtime overhead. With a production rate of more than five MB/s, these generators could seamlessly support a stream cipher, if they were cryptographically secure. Unfortunately they are not, but showed some weaknesses in statistical tests as discussed in the previous section.  Xorshift clearly has the highest throughput. Its derivatives Xorshift64+ and Xoroshiro128+ perform around 2--3 times slower.

From the complex generators,  SHA256PRNG and the Mersenne Twister are presented with additional maximal values in processing time. For SHA256PRNG this is due to the algorithmic property of
producing 32\,Bytes in one hash call, which thereafter are split into 8\,Byte long integer values. Hence, a
hash value is computed only every fifth call, which then takes significantly longer.
The Mersenne Twister follows a similar approach, although it does not involve cryptographic functions. At the
initialization and every 624 calls, it generates 624 fresh pseudo-random integers, which
takes up to 190\,$\mu$s. This is orders of magnitudes longer than simply returning a
value from its buffer.  This notably degrades its performance in comparison with the other lightweight generators.

The three cryptographically secure generators Fortuna, CTR PRNG and SHA256PRNG operate more than one order of magnitude slower than the general purpose counterparts.
All algorithms involve cryptographic
functions (AES-128, SHA-256), which are computationally expensive on constrained
microcontrollers.
Clearly, the SHA256PRNG is the most vital among all crpyto-generators presented, requiring
around 70\,$\mu$s on every hash computation, which leads to an average time of 10\,$\mu$s per
integer with number caching, and a rate of almost 400\,kB/s. The more advanced  NIST Hash DRBG
(SHA-256) doubles the time per integer due to additional hash computation in the feedback path. The implementation,
however, does not allow number caching. This overhead is compensated while requesting larger random
blocks or streams because state updates only occur once after each API call, while generating large
outputs requires multiple hashing.

In comparison to SHA256PRNG, the CTR PRNG generates random four times slower in a stream of
numbers, whereas returning one random integer takes between six and thirty times longer. The CTR PRNG on the other hand has an almost constant runtime per integer. It is noteworthy that up to 16\,Bytes can be obtained by the CTR
PRNG without runtime overhead because this generator computes one AES-128 block on every call that delivers 128\,Bit from which four 32\,Bit integers can be created.
The Fortuna operates slowest when requesting continuous random data and it halves the throughput of the CTR PRNG. It has a constant runtime per integer of
less than 90\,$\mu$s, which on the other hand is faster by a factor of five  than the CTR PRNG. It requires approximately 120\,\% of time as the SHA256PRNG when processing a new hash internally (every eighth call), why the rate is less by a factor of almost ten.

The different design choices between SHA256PRNG and CTR PRNG relate to security implications. Holding a precomputed hash
value in RAM as SHA256PRNG does while only one integer is requested may violate security requirements.
In the presence of protected memory, a generator could secure its state and its cached numbers to reduce the attack surface.
Low-cost devices often lack memory protection mechanisms and as such, a design might be favored that avoids keeping sensitive data in memory for longer intervals in order to prevent (i)
manipulation of future output (see Section~\ref{sec:attack}) as well as (ii) predicting future  random numbers (see Section~\ref{sec:reqs}).
In the presence of frequent re-seeding from fresh entropy, the generator state becomes less sensitive to memory attacks. In contrast, the purpose of caching is performance enhancement and random numbers would be simply returned without update after re-seeding. Still, the
performance per integer request of the CTR PRNG would notably benefit from a caching mechanism.

\begin{figure*}
    \centering
    \subfigure[ROM]{{\includegraphics[width=.49\columnwidth]{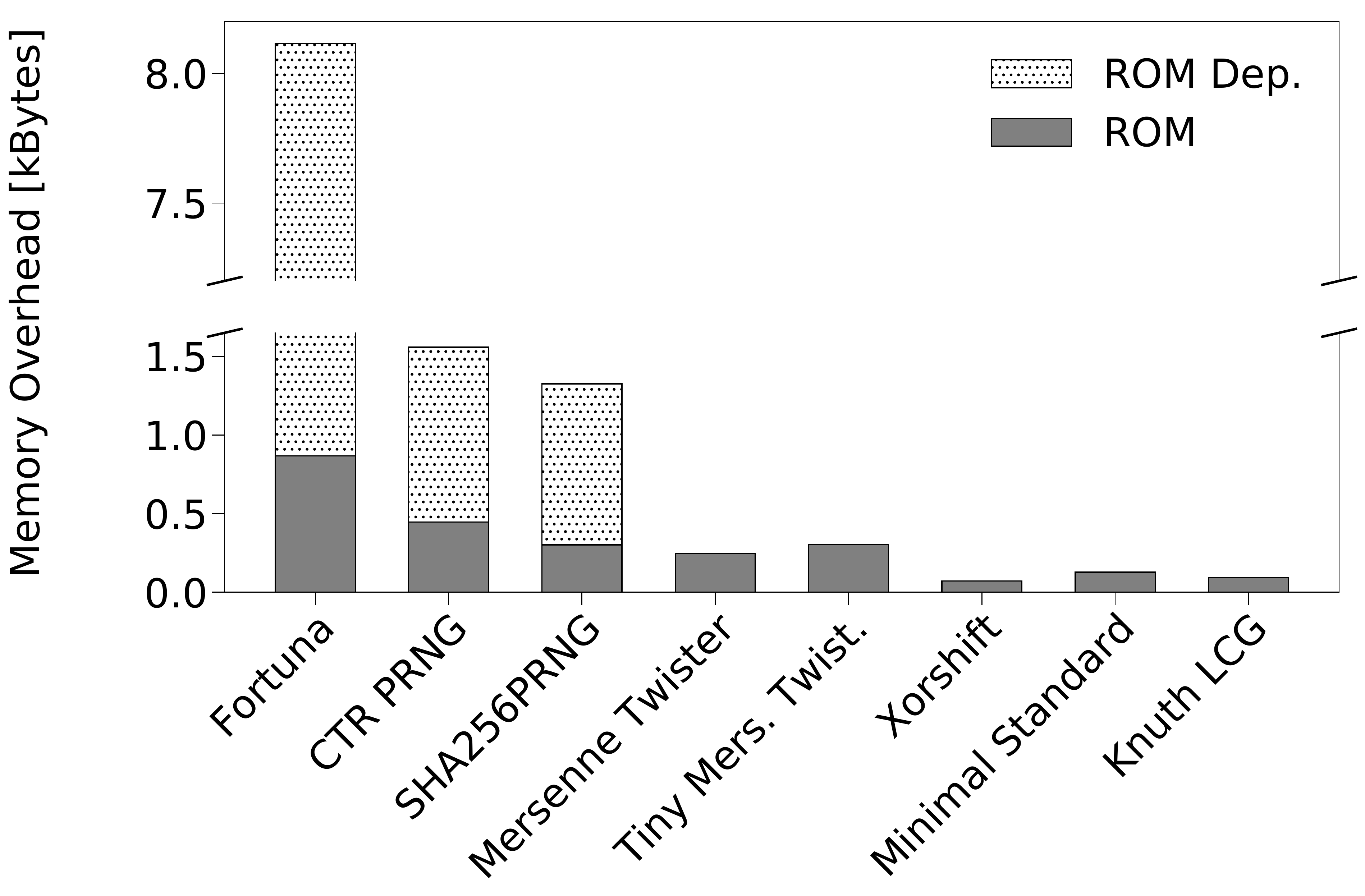} }}
    \subfigure[RAM]{{\includegraphics[width=.49\columnwidth]{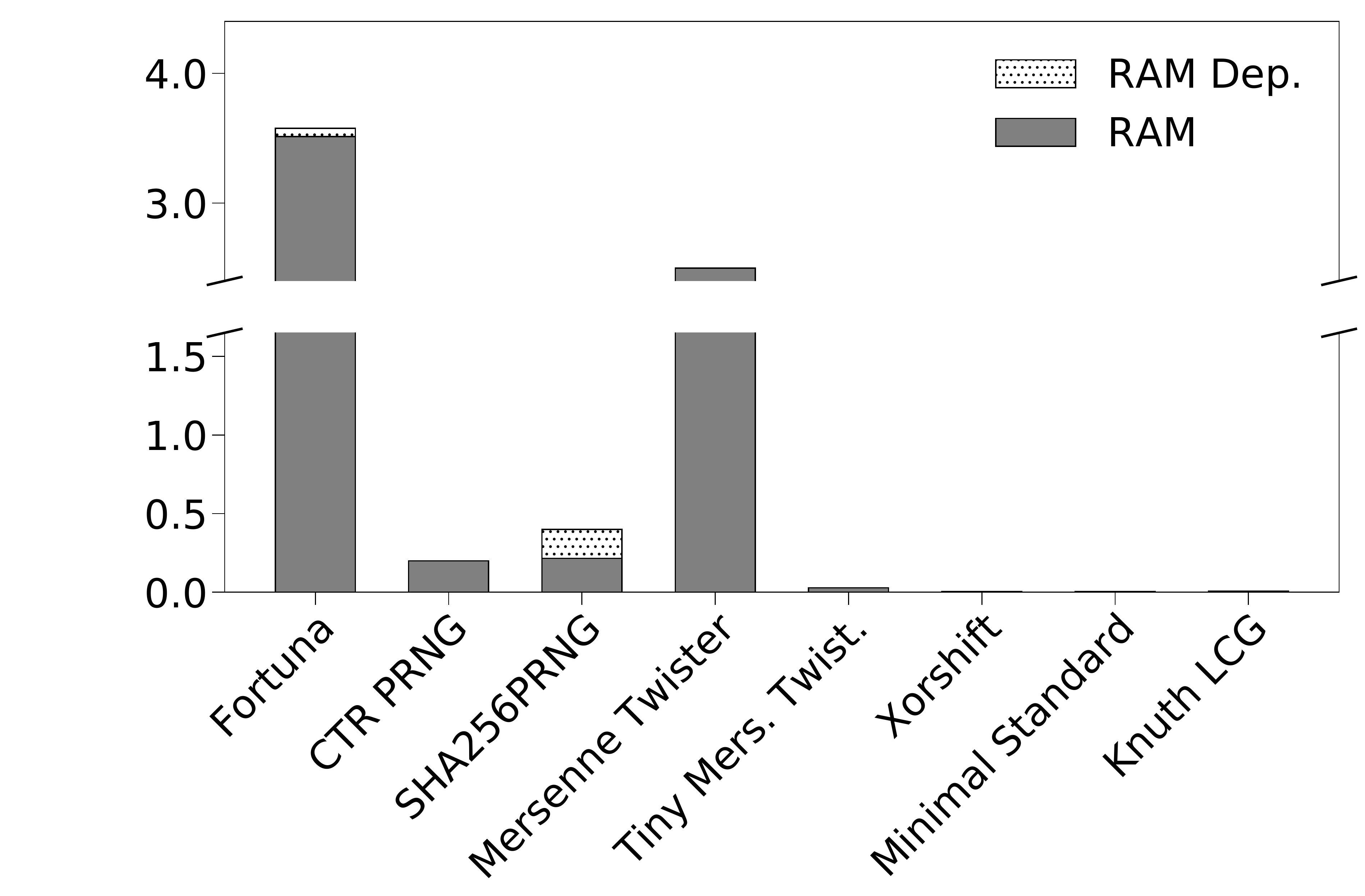} }}
    \caption{PRNG memory overhead in ROM (left) and RAM (right) measured on the STM32F4 microcontroller. ``Dep.'' denotes memory requirements of dependent software modules (\ie hashes, ciphers) that a generator may include.}
    \label{fig:ROM_RAM2}
\end{figure*}

\paragraph{Memory Overhead}
Memory is a particular scarce resource on IoT devices, why memory
of random number generators should have the lowest possible footprint. We evaluate the memory
overhead, which comes on top of a minimal RIOT build while enabling different PRNGs at compile time for the target STM32F4 MCU in Figure~\ref{fig:ROM_RAM2}. Numbers are differentiated w.r.t.  RAM and ROM memory. Furthermore, crypto-purpose generators include dependencies such as hash functions and ciphers, which are highlighted as ``RAM/ROM Dep.''.

Results reassure the unsurprising previous observation that complexity of the PRNGs correlates with resource consumption.
Similar to the throughput measurements, Xorshift, Minimal Standard and Knuth LCG generators remain most
frugal in memory and only require around 100\,Bytes additional ROM and few additional Bytes in
RAM. The internal state size (32 bit), \ie 4\,Bytes in RAM suffice for the Xorshift, whereas the Knuth LCG adds an internal multiplier to a total of 8 Bytes RAM. Similarly, Xorshift64* has an 8\,Byte  state, but it requires additional memory to split and buffer 8\,Bytes generated on each cycle into two 32 bit integers.
In total, Xorshift64* allocates 20 Bytes RAM and Xoroshiro128+ allocates 28 Bytes.
 The Xorshift-derivatives hence operate on a different scale than the ultra-lightweight alternatives.
Both Mersenne Twisters allocate between 250--300\,Bytes in ROM, whereby the
``tiny'' version surprisingly requires the higher amount. The Mersenne Twister is comparably
RAM intensive by allocating up to 2.5\,kB, which is mainly used for its buffer of 624 integers.

Crypto-secure PRNGs such as SHA256PRNG and CTR PRNG are rather efficient in RAM when compared to the Mersenne Twister, but are more demanding in ROM memory. This is mainly due to its dependent cryptographic functions. The NIST Hash DRBG (SHA-256) uses  RAM similar to the SHA256PRNG but it requires additional 800\,Bytes ROM. The reasons relate to its extra logic in the state update function as well as its re-seeding capabilities.
The most demanding generator in terms of ROM memory is the Fortuna due to its dependencies to multiple cryptographic functions. It
includes the SHA-256 and the AES-128 RIOT modules, which is the  main source of high ROM requirement. Furthermore, the
Fortuna internally implements an entropy pool that by default initializes 32 SHA-256 contexts, one of which
requires 104\,Bytes. This is the main reason for its high RAM consumption of around 3,5\,kB. Here it must be stressed
 that this already consumes 10\,\% of the available memory on the test device, or 44\,\% on an Arduino Mega 2560.

\paragraph{Energy Consumption}
We evaluate the energy consumption with the same setup as described in Section~\ref{sec:eval_hw}. Figure~\ref{fig:prng_energy_boxes} displays the energy consumed by each pseudo-random generators in RIOT, measured on the STM32F4 platform. The Fortuna, CTR PRNG and the SHA256PRNG crypto-purpose generators are most expensive in energy---up to several micro Joule per integer. Thereby, SHA256PRNG operates at the lower end and the SHA256-based NIST Hash DRBG consumes about twice as much. The CTR PRNG notably demands maximal resources. Using approximately 24\,$\mu$J it drains a factor of 3.7 more energy than the Fortuna. The CTR PRNG generates four integers in one call, why it outperforms the Fortuna slightly in a stream of random numbers.

All other PRNGs operate on the same scale as the on-chip TRNG on that device (compare Figure~\ref{fig:hwrng_energy_bars}), which takes approximately 0.03\,$\mu$J. Non-cryptographic generators remain by a factor of 20--100 below the most efficient cryptographic generator SHA256PRNG. The most resource-friendly general-purpose PRNGs are the Xorshift with an average of 0.01\,$\mu$J and the Knuth LCG with 0.02\,$\mu$J per integer, which is in agreement with our findings from throughput measurements in Table~\ref{tab:prng_rates}. The enhanced Xorshift64* and Xoroshiro128+ consume 3--4 times more energy than their lightweight 32 bit Xorshift companion, which is a notable increase over the Knuth LCG.

\subsection{Recommendations on PRNGs}

Having examined eleven widely available software-PRNGs  (eight algorithms plus three variants) that cover the basic levels and functions, we are now ready to make the choice of recommendation for random number generators to be included in a constrained IoT operating system. 
Following the previous discussion in Section~\ref{sec:environ}, we differentiate our selections in one general purpose and one crypto-secure PRNG as two different system functions.

\paragraph{General Purpose Generator}
The general purpose generator should be very lightweight, while complying with common statistical requirements. It should run seamlessly on very constrained 8 bit microcontrollers at low energy costs.

From the two ultra-lightweight generators, only the Knuth LCG passes all NIST and DIEHARDER statistical tests. It performs second best. The Xorshift generator consumes about half of its resources, but  has notable statistical flaws. The Knuth LCG produces better output sequences while being similarly fast. The enhanced Xorshift64* or Xoroshiro128+ generators  exceed resource consumption of the Knuth LCG significantly. All other lightweight generators admit lower statistical quality at higher cost. 
We therefore recommend the Knuth LCG to be used as general-purpose, non-cryptographic PRNG in the constrained IoT.

\paragraph{Crypto-secure Generator}
The Fortuna, SHA256PRNG, (NIST Hash DRBG, ) and CTR PRNG are the only
candidates for a CSPRNG, as they are constructed from secure and non-invertible cryptographic functions. Table~\ref{tbl:results-summary} summarizes the security properties and performance and illustrates the design trade-offs as a basis of our recommendation.

\begin{table}
  \caption{Summary of the security properties vers. performance trade-offs for CSPRNGs. Backward secrecy can be enabled with external entropy source (\bpmark). Resource consumption is high ( $\uparrow $), medium ($\rightarrow$), or low ($\downarrow$).}
  \label{tbl:results-summary}
  \begin{tabular}{lccccccc}
  \toprule
  & \multicolumn{3}{c}{Security Properties} & \multicolumn{4}{c}{Performance}
  \\
  \cmidrule(lr){2-4} \cmidrule(lr){5-8}
  Generator & Statistic & Forward & Backward & Runtime & ROM & RAM & Energy
  \\
  \midrule
	  Fortuna & \cmark &  \cmark & \bpmark &  $\rightarrow$ & $\uparrow $ & $\uparrow $ &  $\rightarrow$ 
  \\
	  CTR PRNG &  \cmark &  \cmark & \bpmark &  $\uparrow $ &  $\rightarrow$ & $\rightarrow$ &  $\uparrow $ 
  \\
	  SHA256PRNG &  \cmark &  \cmark\textsuperscript{a} & \bpmark & $\downarrow$ &  $\rightarrow$ & $\rightarrow$ &  $\downarrow$ 
  \\
  \bottomrule
  \end{tabular}
	
    \smallskip
    \footnotesize{\textsuperscript{a} Previous-to-last random value not protected by forward secrecy.}
\end{table}

The CRPRNGs pass all statistical tests (including BigCrush) and fulfill the requirements of unpredictability and brute-force resistance. Perfect forward and backward secrecy after a state compromise is assured by
the Fortuna, CTR PRNG and the NIST Hash DRBG based on their one-way nature during state update in 
combination with re-seeding. The SHA256PRNG holds a (linear combination of) the previous-to-last state and thus only guarantees forward secrecy for earlier values due to its secure one-way SHA-256 output function (cf.,~\cite{kswh-capng-98}). Backward secrecy can be added by re-seeding on demand to all generators as discussed in~\ref{sec:reseed_csprng}  using predefined interfaces to entropy sources. 

 The SHA-256 generator stands out as it is the only cryptographically secure algorithm that attains moderate performance values. All competitors exceed the SHA-256 performance measures by one order of magnitude in  at least one dimension. We conclude that the resource frugality in combination with modular cryptographic robustness justify to commend SHA256PRNG as the CSPRNG in the constrained IoT.

\section{Random Numbers on AI Platforms}\label{sec:ai}

Machine learning and other algorithms of Artificial Intelligence (AI) recently gained attention and are considered for deployment at the IoT Edge. They shall serve use cases such as voice recognition, object counting, or anomaly detection.
Machine learning, in particular reinforcement learning makes heavy use of random numbers, since randomized algorithms (\eg Monte-Carlo methods~\cite{p-mcmic-78}) are involved to explore state spaces.
Software libraries for machine learning evolve and exist already, optimized for constrained embedded devices~\cite{ddjrj-tlmem-20, brlff-btscd-20}.

High processing demands triggered a new set of hardware platforms to facilitate computation of AI algorithms at minimal energy consumption.
A common approach is represented by dual core SoCs and hardware accelerators to offload the main CPU from processing.
We identified the following three categories in the data sheets: signal processors, neural network accelerators, and tailored hardware engines for assisted audio-video processing.
Commercial products largely base on ARM Cortex-M processors with a proprietary instruction set architecture (ISA). The academic community introduced a series of open source RISC-V processors that vary with dedicated purposes while targeting low energy consumption~\cite{scrgp-sswrc-17}.
Parallel ultra low power (PULP) processors feature an optimized processor design. Multiple cores can be clustered and share a  coarse-grained  memory architecture, the instruction cache and peripherals. Energy efficiency is achieved by speedup after parallelization and operating the cores `near threshold' by applying voltage and frequency gating~\cite{gstlp-ntrcd-17}. This extends utility in the contexts of signal-, neural network-, or audio-video processors. PULP clusters can be augmented~\cite{csspr-iesse-17} by  cryptographic hardware engines to facilitate low-overhead encryption of network traffic, or by convolution engines.

The modular design and performance of these novel RISC-V based processors have been analyzed and simulated by De Giovanni~\etal~\cite{gmdmq-mdoba-20}. They observe a speedup factor of five  and energy savings of 40\,\% for an 8-core PULP cluster over a single core. Additional hardware acceleration decreases energy consumption down to 50\,\%.
Their findings are in agreement with
Wang~\etal~\cite{wmcb-foten-20}, who analyzed neural network inference on constrained IoT devices. They compare the processing overhead and energy consumption of a machine learning algorithm running on an off-the-shelf ARM Cortex-M4 node and two optimized RISC-V platforms, one single-core (RI5CY) and one 8-core cluster of RI5CY processors (\textit{Mr. Wolf} IoT processor~\cite{prlmb-mwgep-18} for high processing demands).
Their results indicate that RI5CY outperforms Cortex-M by a factor of up to 1.3 for fixed point integer arithmetics. The neural network use case provides a speedup of six times and reduces the energy consumption by up to 70\,\%  on the multi-core platform in comparison with a single-core Cortex-M4 operation.

The shift of computational complexity towards edge devices imposes new requirements on the random number generation. The  rate of statistical (pseudo-)random numbers consumed at low-end edge devices increases, whereas crypto-secure random generation remains unaffected by this paradigmatic shift.
Conversely, new AI platform architectures may counter these increased demands.

We evaluate different pseudo-random number generators (cf.,~\autoref{sec:eval_prng}) on two single-core RISC-V based processors of the PULP family and compare the performance to an off-the-shelf ARM Cortex-M core.
\autoref{tbl:risc_hw_compare} summarizes the hardware properties of our reference platform: The VEGAboard~\cite{openisa-vegaboard} holds multiple cores on the RV32M1 chip, which it can operate independently. We use this platform to compare ARM and RISC-V properties. In terms of processor complexity and application targets, the ZERO-RISCY~\cite{pulp-zero-riscy-18} is on par with a Cortex-M0+ processor, while RI5CY~\cite{pulp-ri5cy-19} is on par with Cortex-M4. All four cores operate at the same CPU frequency and share memory as well as peripherals which includes a TRNG.

\begin{table}
    \caption{Overview of the different processor cores on the RV32M1 microcontroller. It runs at 48\,MHz and provides 384\,kB RAM, 1.2\,MB internal and 4.0\,MB external flash memory. The naming scheme IEFCM resolves to the following architectural components. I: base integer instruction set, E: embedded base integer instruction set, F: single precision floating point extension, C: extensions for compressed instructions, M: integer multiplication and division extension.}
    \label{tbl:risc_hw_compare}
    \setlength{\tabcolsep}{4pt}
    \begin{tabular}{c l c l}
            \toprule
            \makecell[c]{Processor Core}&
            \makecell[c]{Instr. Set\\Architecture}&
            \makecell[c]{Pipelining\\$[$\# stages$]$}&
            \makecell[c]{Special Features}\\
            \midrule
            \makecell[r]{ZERO-RISCY}&RV32-IECM\textsuperscript{a}&2&\makecell[l]{Area optimized (2.2\,x smaller than RI5CY)\\Energy boost for mixed control/arithmetic code}\\
            \midrule
            \makecell[r]{RI5CY}&RV32-I(F)CM\textsuperscript{b}&4&\makecell[l]{Post-incrementing load and stores\\Single-cycle multiply-add \& ALU extensions\\Auto increment hardware loops\\Memory protection unit}\\
            \midrule
            ARM Cortex-M4F&ARMv7E-M\textsuperscript{c}&3&\makecell[l]{Branch speculation engine\\Single-cycle multiply-add extension\\Memory protection unit}\\
            \bottomrule
        \end{tabular}

\end{table}

\autoref{tbl:ai_prng_rates} presents our measurement results running five  non-crypto PRNGs on three cores of the VEGAboard platform, as well as the embedded TRNG.
The pseudo-random number throughput ranges from 1000--6000\,kB/s on ZERO-RISCY, and RI5CY speeds up by a factor of 1.1--1.4 which is achieved by additional hardware extensions of the processor.
The Cortex-M4 increases performance of PRNGs that involve multiple XOR operations (Xorshift, Mersenne Twister), though, RI5CY outperforms the Cortex-M4 by 10\,\% for the others, making effect of the ALU extensions and hardware loops.

Energy demands reflect similar properties with $\approx$ 20\,nJ consumption on both RI5CY and Cortex M4 while operating the recommended Knuth LCG. The fastest PRNG Xorshift is content with 10\,nJ per integer on the Cortex-M4, which is only 75\,\% of the RI5CY consumption.
The TRNG performance  operates at the lower end compared to other hardware generated random numbers (\autoref{subsec:hwperform}), however, it should be involved only for seeding.

Despite small variations, the performance of random number generation on a single core scales similarly to our measurements presented in~\autoref{subsec:eval_prng_performance}, regardless of the processor architecture. Our measurements are also in rough agreement with the single-core results presented by Wang~\etal~\cite{wmcb-foten-20} who found a speedup factor of 1.3 for RI5CY over Cortex-M, operating neural networks. For multi-core processors, we expect the same upscale to hold and $\approx$ 6\,x speedup with about 50--70\,\% energy savings for pseudo-random number generation.
In the context of machine learning at the edge, random input can be parallelized with these concurrent hardware architectures, which addresses high processing requirements at low energy consumption.

As an alternative, Forooghifar~\etal~\cite{faa-rdemu-19} introduce ``self-awareness'' as an architecture-independent solution to improve nodal lifetime when operating machine learning at the edge. Complex computations are outsourced from battery driven nodes and distributed between the edge, fog, and cloud, following an energy estimation on the constrained node.

Processing demands for AI are and will continue to be costly on IoT edge devices.
In contrast, many simple sensor nodes will commonly process only little data and require only a single processor. Consequently, performance enhancements of a targeted AI hardware platform will enfold limited impact in conventional IoT use cases, as indicated by our measurement results.

\begin{table}[]
    \caption{Throughput and energy consumption per integer (\#) for non-crypto PRNGs running on three different processor cores of the VEGAboard.}
    \label{tbl:ai_prng_rates}
    \centering
    \begin{tabular}{ l r r r   r r r}
        \toprule
        & \multicolumn{3}{c}{Byte rate $[$kB/s$]$} & \multicolumn{3}{c}{Average energy per \# $[$nJ$]$\textsuperscript{a}}\\
        \cmidrule(lr){2-4}
        \cmidrule(lr){5-7}
        Generator & \makecell{ZERO-RISCY} & \makecell{RI5CY} & \makecell{Cortex-M4} & \makecell{ZERO-RISCY} & \makecell{RI5CY} & \makecell{Cortex-M4}  \\
        \midrule
        {Mersenne Twister} &  1118  & 1527 & 2290 & 79.5&60.3&43.3\\
        {Tiny Mers. Twist.}&  2158  & 2935 & 2608 & 49.3&36.7&39.4\\
        {Xorshift}         &  5869  & 6264 & 7227 & 14.8&13.2&10.9\\
        {Minimal Standard} &  1715  & 2331 & 2134 & 61.9&46.9&50.9\\
        {Knuth LCG}        &  3682  & 4697 & 4581 & 26.3&20.2&20.2\\
        \cmidrule(lr){2-4}
        \cmidrule(lr){5-7}
        {TRNG}             &  \multicolumn{3}{c}{2.4$\cdot 10^{-2}$}&\multicolumn{3}{c}{2.1$\cdot 10^{6}$}\\
        \bottomrule
        \end{tabular}

        \smallskip
        \begin{flushleft}
        \footnotesize{\textsuperscript{a}Standard deviations are low for PRNGs ($\sigma$ $<$\,1.5\,\%) and high for the TNRG ($\sigma$ $\approx$ 22\,\%).}
        \end{flushleft}
\end{table}

\section{Discussion: Hardware or Software for Randomness in the~IoT?}\label{sec:hwrng_vs_prng}

An increasing number of embedded controllers is expected to provide hardware primitives for basic cryptographic operations in the near future, which bring promise of  fast and efficient random generators and contribute real entropy. The performance of on-chip hardware based random number generators, however, is heterogeneous. Some devices operate fast and save battery resources (\ie STM32F4), while others
are slow and require notably larger amounts of energy (\ie nRF52840) than corresponding software. As shown in Section~\ref{sec:eval_hw}, few devices even produce poor statistical output (\ie CC2538). Some manufacturers advice against an immediate use for cryptographic random number.  Other manufacturers describe their on-chip random number generators as suitable for cryptographic purposes without providing a cryptographic proof (Nordic).
Truly random generators rely on real entropy from a physical process. This may be influenced by environmental factors, which opens an attack surface.

Common IoT operating system such as RIOT need to make decisions on which hardware functions to include and how to integrate hardware and software components into the random subsystem. These multi-platform  multi-purpose systems want to provide   an overall lean solution of reliable quality at a predictable performance. To aid this design process, we now analyze key performance properties of the different hardware- and software-based random number generators that are either unseeded (TRNG), seeded ((CS)PRNG), or hybrid.

\begin{figure*}
    \centering
    \includegraphics[width=1\columnwidth]{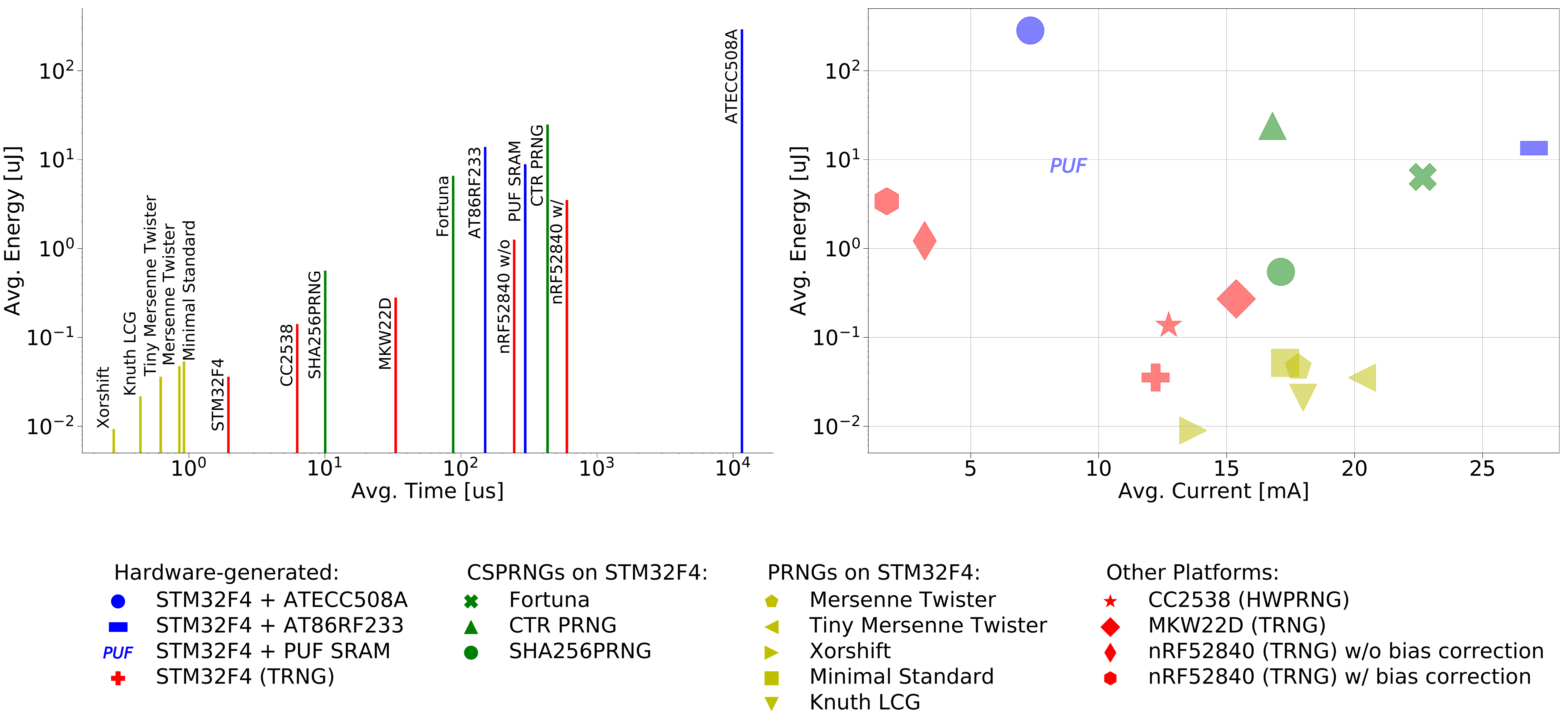}
    \caption{{Average energy consumption over average time (left) and average current draw (right) for hardware and software generated random integers.}}
    \label{fig:hwrng_energy_stm32_scatter}
\end{figure*}

In Figure~\ref{fig:hwrng_energy_stm32_scatter}, we compare energy consumption and average run-time per integer (left)  as well as the average current flow (right) for the hardware systems summarized in Table~\ref{tab:hw_overview}. Software PRNGs are measured on the STM32F4 board.
We observe that the lightweight software generators are fastest and consume the least energy together with the TRNG on the chip of STM32F4 . While
software PRNGs operate up to five times faster, they charge a higher current than the TRNG which results in a similar energy
footprint. On-chip hardware generators on the CC2328 (HWPRNG) and MKW22D (TRNG) devices consume about one order of magnitude more time and energy. Hence, we argue for the choice of a lightweight software generator (\ie the Knuth LCG) for the regular production of random numbers. Hardware generators are best used for (re-)seeding, when true entropy is required. 

Running a
cryptographically secure generator demands unsurprisingly more energy resources in comparison to its lightweight PRNG
alternatives or the STM32F4 on-chip TRNG.
Among the three software CSPRNGs, SHA256PRNG is clearly  most frugal with an energy consumption of 0.5\,$\mu$J for one integer and an average current consumption of 17\,mA.
Note that caching is involved here, thus, only every eighth requested integer involves computation of a new hash which then causes currents up to 20\,mA. Fortuna and
CTR PRNG drain notably more energy in comparison to the SHA256PRNG as already visible in the previous analysis. CTR PRNG consumes up to 24\,$\mu$J for one integer, which is four times more than 
Fortuna, even though its average current consumption is similar to the SHA256PRNG that uses 5\,mA less than  Fortuna. The increased energy consumption for one integer may be compensated
in a stream or by caching intermediate numbers.
Hardware based random numbers on the nRF52840 controller
consume energy similar to the CSPRNGs, although the low-power MCU drains less than 5\,mA in both operation modes (w/ and w/o bias correction). This is due to its slow operation.

The external ATECC508A chip, which implements an approved
HWCSPRNG with seeding in hardware, exhibits a worse performance than all software CSPRNGs. The throughput of the software solutions is 20--100 times higher while the energy consumption
remains 3--10 times less.
The average current of the ATECC508A solution remains small in comparison to the software solutions that do not have to power a second device. This is because the main controller idles while the external chip processes random values. The external chip requires only a small current as shown in Section~\ref{subsec:hwperform}. Eventually, the total energy consumption can be improved by further driver optimizations. A minimum of 280\,$\mu$J is required to request from one up to eight integers.
It is noteworthy that this chip implements secure seeding on its own, which enhances very constrained devices without proper entropy sources. Furthermore, it implements tamper detection in hardware which provides additional protection against side channel attacks with physical device access.

Memory limitations of very constrained devices (\eg ATmega2560 with only 8\,kB RAM and 256\,kB Flash) are unable to run
 complex software CSPRNGs due to memory constraints. The Fortuna CSPRNG requires almost 50\% of the ATmega2560 RAM leaving 
only 4\,kB for firmware and the remaining security protocols, which is insufficient for real-world IoT-networking applications~\cite{bghkl-rosos-18}.
Instead, the
device driver for the ATECC508A chip can be included which (i) reduces  memory requirements of cryptographic functions, (ii) offloads processing
of complex algorithms on the device, and (iii) includes additional crypto-related features.

Next to the dedicated crypto-chip, other external randomness generators, namely the AT86RF233 transceiver as well as the SRAM PUF
 are  energy expensive using up to one order of magnitude more than the SHA256PRNG. Both mechanisms, though, are not designed to be used periodically. Instead, they act as seed sources that are utilized once
during instantiation of a PRNG and on re-seeding. The current consumption differs notably between the transceiver and the PUF
SRAM. The transceiver powers two devices in active mode because the MCU polls random bytes via the SPI while
the radio needs to stay in receive mode. This draws a current of up to 26\,mA in total. The PUF SRAM on the other hand takes
twice as long to retrieve a high entropy integer but it drains less than 10\,mA on average. As depicted in
Section~\ref{subsec:puf}, this procedure needs to take place very early on startup---even before clock and bus
initialization in the operating system.

In summary, we argue that seeded pseudo-random number generators---a lightweight general purpose PRNG or an approved CSPRNG---are the preferable solution for producing (secure) random numbers. Co-processors and external hardware assistance are vital for adding entropy  and can
help to reduce memory footprints for tiny devices. They lead to a decreased throughput, though, when compared to software that runs on the main controller. Conversely, special crypto-chips can offload processing demands from the main controller. Transceivers as well as uninitialized SRAM are essential on devices lacking TRNGs.
Even though PUF SRAM and the radio-based entropy sources are costly, they uniquely contribute entropy and are rarely needed  for seeding purposes.

\section{Conclusions and Outlook}\label{sec:outro}

In this paper, we explored the building blocks for randomness  in the constrained Internet of Things: hardware and software components that generate statistical randomness, entropy, and resilience against cryptographic attacks. We systematically derived the requirements for IoT random subsystems from the perspectives of statistics, security, and operating system integration. An extensive, comparative evaluation using several prominent test suites as well as detailed performance measurements on popular devices delivered insights into the overall quality and suitability of the different components under test.
This work derives four major recommendations:

\begin{enumerate}
\item Separate general purpose random generators from cryptographically secure generators on the OS level. Avoid any mixture or interference between the two.
\item Prefer (software) PRNGs over random hardware, as they are more efficient and reliable. Exploit hardware components as additional entropy sources for (re-)see\-ding or when CSPRNG operation is infeasible on a constrained node.
\item The Knuth LCG  is the most efficient general purpose generator that provides decent statistical quality. It is simple and lean enough to run on very constrained devices.
\item We recommend SHA256PRNG as a cryptographically secure generator, since it outperforms its competitors by an order of magnitude in several dimensions.
\end{enumerate}

With this work, we hope to contribute to a thoughtful development toward a secure Internet of Things. This will be of particular importance, as more and more (sensitive) data originates from IoT nodes and needs protection. Content object security with OSCORE~\cite{RFC-8613,gasw-icoso-20} and LAKE~\cite{draft-ietf-lake-reqs}, for example, will facilitate the encryption of individual information units, but will extend the use of cryptographic primitives such as random numbers during operation.

\balance

\bibliographystyle{ACM-Reference-Format}
\bibliography{./local,own,rfcs,ids,theory,complexity,layer2,internet,iot,transport,security,ngi,meta}

\clearpage
\appendix
\newcommand{\mainpaper}[1]{the main article [#1]}

\section{Hardware Generated Random Numbers}\label{appendix:eval_hw}
\subsection{Statistical Analysis with NIST STS}\label{appendix:hwsts}
\begin{figure*}
  \centering
  \includegraphics[width=\textwidth]{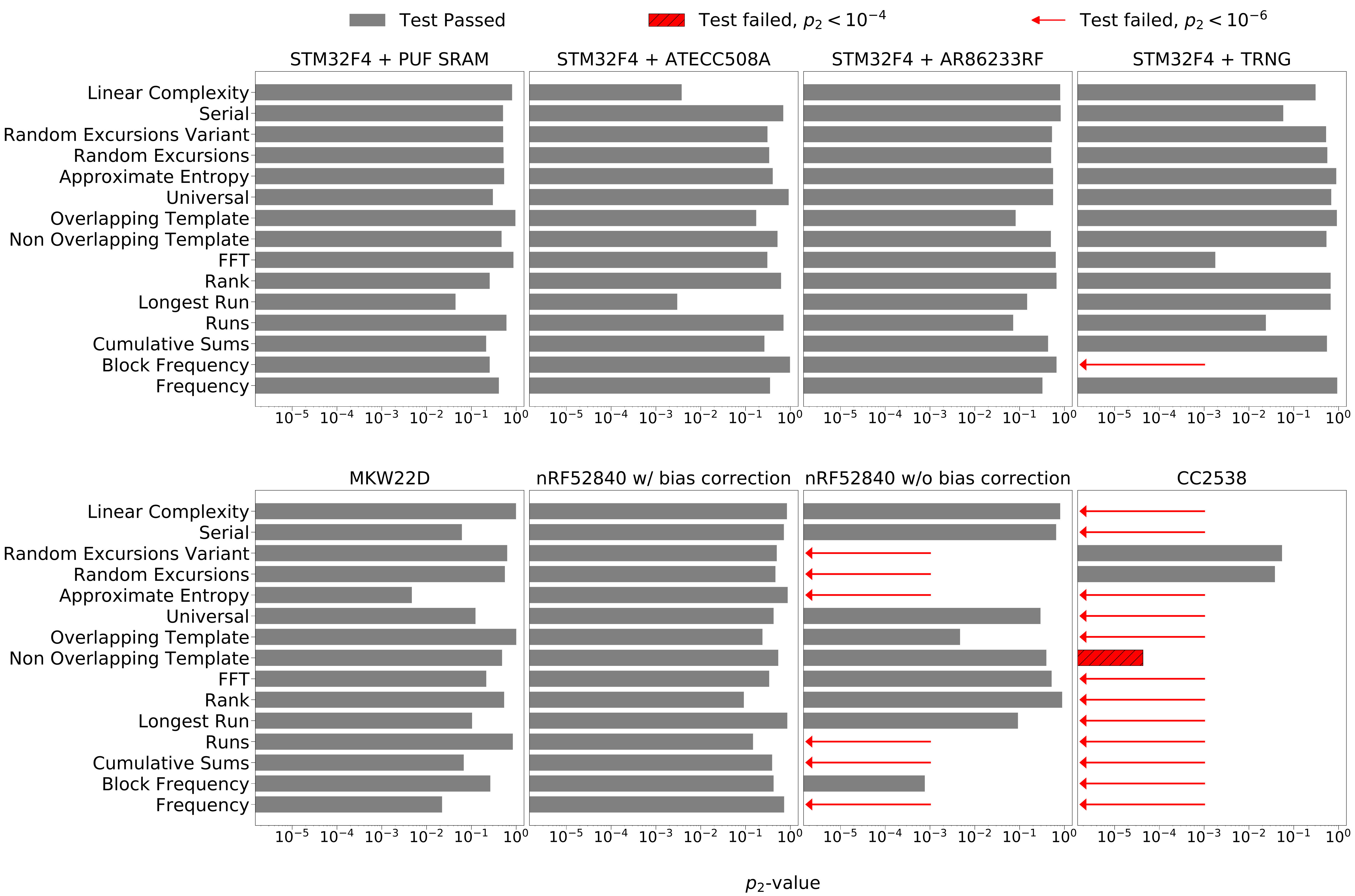}
  \caption{Hardware Generated Random Numbers: $\chi^2$-test results on the distribution of probability values from 15 NIST STS tests. $p_2$-values $\geq$ 0.0001 pass the hypothesis of uniformity.}
  \label{fig:HW_nist_pval_subplots}
\end{figure*}

Results of the $\chi^2$-test on the distribution of $p$-values are shown in Figure~\ref{fig:HW_nist_pval_subplots}. 
Certain tests produce multiple $p_2$-values and we show the average in that case.
We display passing tests with gray bars, and we highlight failed tests
The test suite additionally calculates a proportion of passed test runs and evaluates its significance. Passing tests in Figure~\ref{fig:HW_nist_pval_subplots}
have a significant proportion within the confidence interval calculated from $\alpha=0.01$.
with hatched red bars or red arrows in case that the $p_2$-value is too small to be displayed.
Details are discussed in \mainpaper{Section~6.2}.

\clearpage
\section{Software Generated Pseudo-Random Numbers}\label{appendix:eval_prng}
\subsection{Statistical Analysis with NIST STS}\label{appendix:prngsts}
\begin{figure*}
  \centering
  \includegraphics[width=\textwidth]{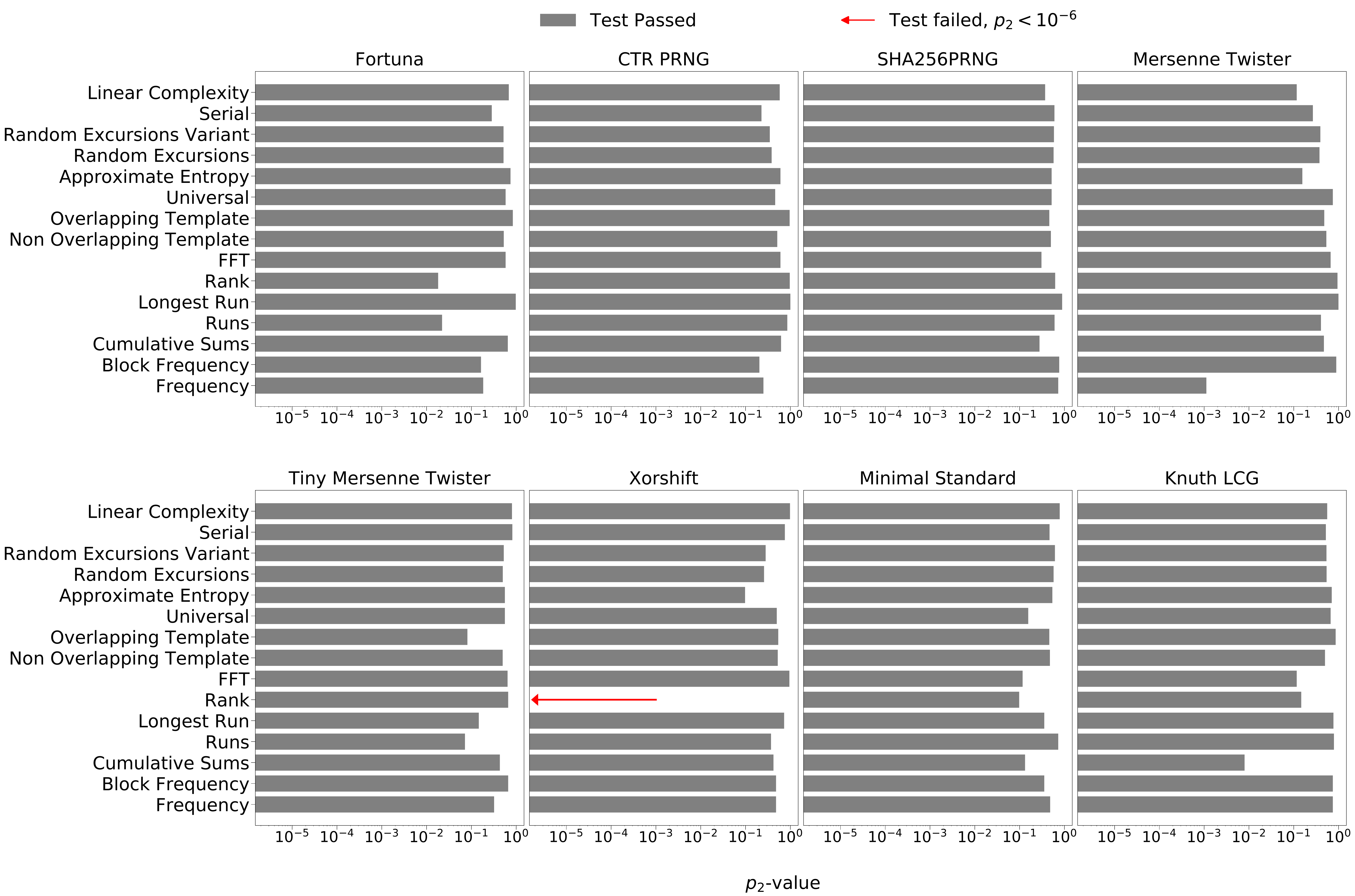}
  \caption{Pseudo Random Numbers: $\chi^2$-test results on the distribution of probability values from 15 NIST STS tests. $p_2$-values$\geq10^{-4}$ pass the hypothesis of uniformity.}
  \label{fig:nist_pval_subplots}
\end{figure*}

The results of the $\chi^2$-test are shown in Figure~\ref{fig:nist_pval_subplots}. 
We display average $p_2$-values, where applicable.
Details are discsused in \mainpaper{Section~7.3}.

\clearpage
\subsection{Statistical Analysis with DIEHARDER}\label{appendix:prngdie}
\begin{figure*}
  \centering
  \includegraphics[width=\textwidth]{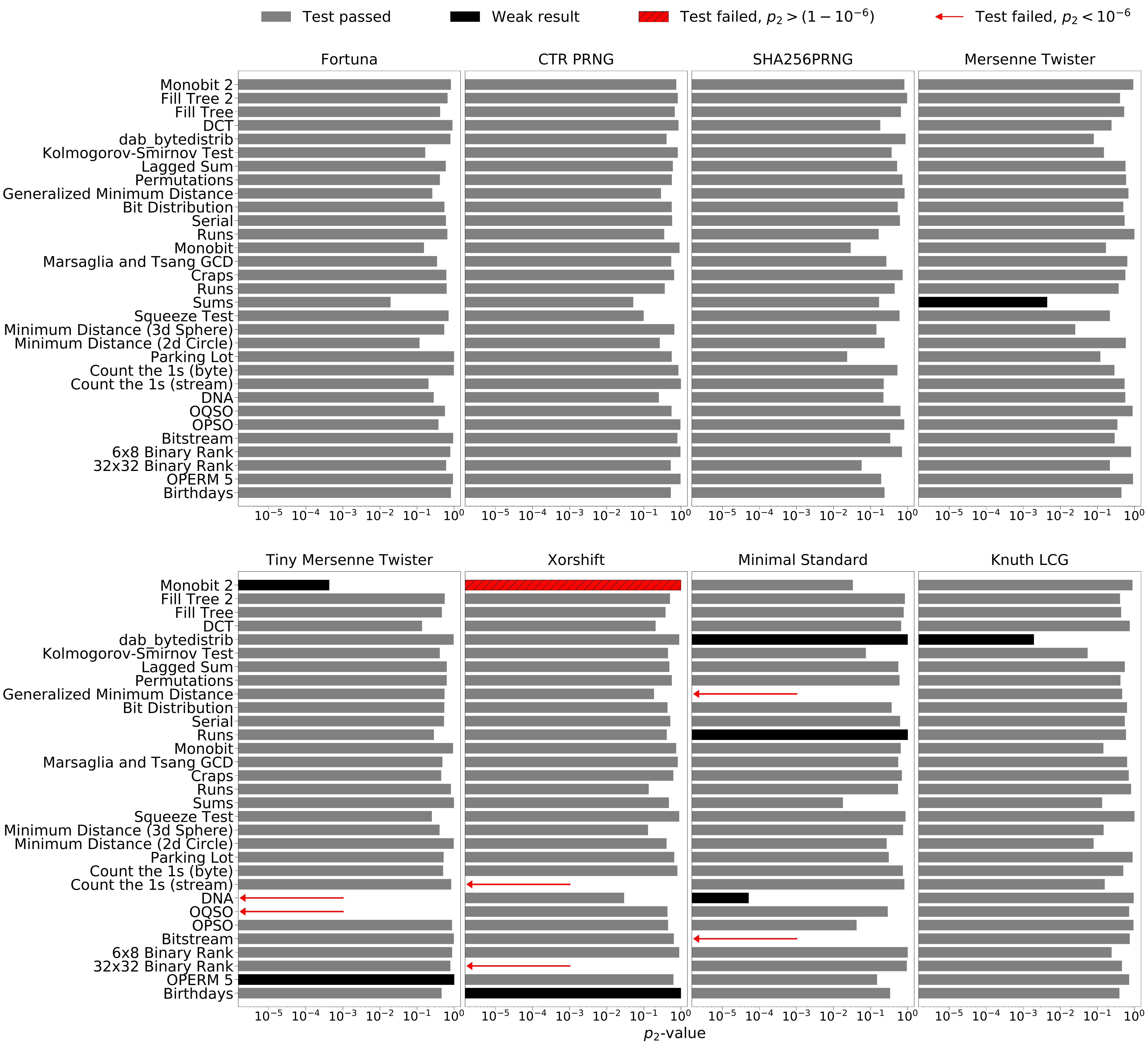}
  \caption{KS-test results on the distribution of probability values from 31 DIEHARDER tests. $p_2$-values in the significance
  interval $(\alpha, 1-\alpha)$ with $\alpha = 10^{-6}$ pass the test. $p_2$-values within the
  interval $(\alpha_2, 1-\alpha_2)$ with $\alpha_2 = 5\cdot10^{-3}$ are considered as weak.}%
  \label{fig:die_pval_subplots}
\end{figure*}

Results of the Kolmogorov-Smirnov test for each PRNG in RIOT are displayed in Figure~\ref{fig:die_pval_subplots}. Similar to the NIST tests, we plot average $p_2$-values, where applicable. In the style of previous graphs, we display passing tests with gray bars, and we highlight failed tests with hatched red bars or red arrows in case that the $p_2$-value is too small to be displayed. Additionally, we mark weak results with black bars.

All complex generators  displayed in the first row of Figure~\ref{fig:die_pval_subplots} pass all tests.
Several failures must be observed for the lightweight generators displayed in the second  row of Figure~\ref{fig:die_pval_subplots}.
Details are discussed in \mainpaper{Section~7.4}.

\clearpage

\section{List of Abbreviations}\label{sec:appendix}

\begin{table}[H]
    \caption{List of abbreviations that we use within this work.}
    \label{tab:abbreviations}
    \centering
    \begin{tabular}{l l}
        \toprule
        \textbf{AES}& Advanced Encryption Standard\\
        \textbf{AI}& Artificial Intelligence\\
        \textbf{ALU}& Arithmetic Logic Unit \\
        \textbf{CPU}& Central Processing Unit\\
        \textbf{CSMA}& Carrier Sense Multiple Access\\
        \textbf{CSPRNG}& Cryptographically Secure Pseudo-Random Number Generator\\
        \textbf{DH}& Diffie-Hellman (key exchange)\\
        \textbf{DRBG}& Deterministic Random Bit Generator\\
        \textbf{DSA}& Digital Signature Algorithm\\
        \textbf{ECC}& Elliptic Curve Cryptography\\
        \textbf{ECDH}& Elliptic Curve Diffie-Hellman (key exchange)\\
        \textbf{ECDSA}& Elliptic Curve Digital Signature Algorithm\\
        \textbf{FIPS}& Federal Information Processing Standard\\
        \textbf{HWCSPRNG}& Hardware Cryptographically Secure Pseudo-Random Number Generator\\
        \textbf{HWPRNG}& Hardware Pseudo-Random Number Generator \\
        \textbf{IoT}& Internet of Things\\
        \textbf{ISA}& Instruction Set Architecture\\
        \textbf{KS-Test}& Kolmogorov Smirnov Test (statistical test)\\
        \textbf{LCG}& Linear Congruential Generator\\
        \textbf{LFSR}& Linear Feedback Shift Register\\
        \textbf{MCU}& Microcontroller\\
        \textbf{M2M}& Machine to Machine\\
        \textbf{NIST}& National Institute of Standards and Technology\\
        \textbf{NIST STS}& NIST Statistical Test Suite\\
        \textbf{OS}& Operating System\\
        \textbf{PULP}& Parallel Ultra Low Power (processor) \\
        \textbf{PRNG}& Pseudo-Random Number Generator\\
        \textbf{PUF}& Physically Unclonable Function\\
        \textbf{RNG}& Random Number Generator\\
        \textbf{RO}& Ring Oscillator\\
        \textbf{RSA}& Rivest-Shamir-Adleman (cryptosystem)\\
        \textbf{SHA}& Secure Hash Algorithm\\
        \textbf{SoC}& System on a Chip\\
        \textbf{TRNG}& True Random Number Generator\\
        \bottomrule
    \end{tabular}
\end{table}

\end{document}